\documentclass[journal]{IEEEtran}

\usepackage{graphicx}
\usepackage{color}
\usepackage{placeins}
\usepackage{float}
\usepackage{tabularx,colortbl}
\usepackage{epstopdf}
\usepackage{amsmath}
\usepackage{amssymb}
\usepackage{caption}
\usepackage{subcaption}
\usepackage{algorithm}
\usepackage{algorithmic}
\usepackage{multirow}
\usepackage{url}

\newtheorem{definition}{Definition}
\newtheorem{Prop}{Proposition}
\newtheorem{Theo}{Theorem}

\begin{document}

\title{Quality-Driven Resource Allocation for Full-Duplex Delay-Constrained Wireless Video Transmissions}

\author{Chuang Ye, M. Cenk Gursoy, and Senem Velipasalar
\thanks{The authors are with the Department of Electrical
Engineering and Computer Science, Syracuse University, Syracuse, NY, 13244
(e-mail: chye@syr.edu, mcgursoy@syr.edu, svelipas@syr.edu).}}

\maketitle

\begin{abstract}
In this paper, wireless video transmission over full-duplex channels under total bandwidth and minimum required quality constraints is studied. In order to provide the desired performance levels to the end-users in real-time video transmissions, quality of service (QoS) requirements such as statistical delay constraints are also considered. Effective capacity (EC) is used as the throughput metric in the presence of such statistical delay constraints since deterministic delay bounds are difficult to guarantee due to the time-varying nature of wireless fading channels. A communication scenario with multiple pairs of users in which different users have different delay requirements is addressed. Following characterizations from the rate-distortion (R-D) theory, a logarithmic model of the quality-rate relation is used for predicting the quality of the reconstructed video in terms of the peak signal-to-noise ratio (PSNR) at the receiver side. Since the optimization problem is not concave or convex, the optimal bandwidth and power allocation policies that maximize the weighted sum video quality subject to total bandwidth, maximum transmission power level and minimum required quality constraints are derived by using monotonic optimization (MO) theory.
\end{abstract}
\thispagestyle{empty}


%
\IEEEpeerreviewmaketitle

\section{Introduction}
Recently, with rapid developments in communication technology, multimedia applications such as video telephony, teleconferencing, and video streaming which are delay sensitive and bandwidth intensive, have started becoming predominant in data transmission over wireless networks. For instance, as revealed in \cite{Cisco}, mobile video traffic accounted for 60\% of the total mobile data traffic in 2016, and more than three-fourths of the global mobile data traffic is expected to be video traffic by 2021. Indeed, mobile video has the highest growth rate of any application category measured among the mobile data traffic types. Such dramatic increase in wireless video traffic, coupled with the limited spectrum resources, brings a great challenge to today's wireless networks. Therefore, it is important to improve the wireless network capacity by allocating the limited resource efficiently. In such multimedia applications, certain quality of service (QoS) guarantees also need to be provided in order to satisfy the performance requirements of the end-users. For instance, in order to ensure a satisfactory user experience, bounds on time delay are imposed in real-time video transmissions. The strictness of the delay constraints varies based on the specific wireless multimedia application. For instance, live video streaming may tolerate some delay whereas bidirectional video conferencing requires much more stringent time delay bounds on the order of few milliseconds in order to guarantee satisfactory user experience. Supporting such QoS requirements with stringent delay limitations requires larger transmission rates that can be achieved by using more resources such as bandwidth and power, and facing less interference. Therefore, it is critical to allocate the limited resources efficiently taking into account the QoS requirements of different users in the wireless network.

The authors in \cite{Khalek} proposed a strategy to maximize the sum quality of the received reconstructed videos subject to different delay constraints on different users and a total bandwidth constraint in a multiuser setup by allocating the optimal amount of bandwidth to each user in a downlink wireless network. They also derived user admission and scheduling policies that enable selecting a maximal user subset such that all selected users can meet their statistical delay requirements. A content-aware framework for spectrum- and energy-efficient mobile association and resource allocation in wireless heterogeneous networks was proposed in \cite{Yiran}. Two content-aware performance metrics, namely quality-of-experience-aware spectral efficiency (QSE) and quality-of-experience-aware energy efficiency (QEE), were used to capture spectrum usage and energy consumption from the perspective of video quality. The goal was to obtain the optimal system level QSE and QEE by determining the mobile association and allocating the resources optimally via nonlinear fractional programming approach and dual decomposition method. In this work, delay QoS constraints were not considered. On the other hand, reference \cite{Ma} addressed the maximization of the system throughput subject to delay QoS and average power constraints for time-division multiple access (TDMA) communication links. \cite{Jia} proposed a QoS-driven power and rate adaptation scheme that aims at maximizing the throughput of multichannel systems subject to a given delay QoS constraint over wireless links. Multichannel communication can achieve high throughput and satisfy stringent QoS requirements simultaneously. The authors in \cite{Yichen} developed an optimal power allocation scheme for the cognitive network with the goal of maximizing the effective capacity of the secondary user link under constraints on the primary user's outage probability and secondary user's average and peak transmission power. The scheme also satisfied the QoS requirements of both secondary users and primary users simultaneously. Statistical QoS provisioning in next generation heterogeneous mobile cellular networks was investigated in \cite{Alireza}. Under certain assumptions, a lower bound for the system performance was introduced in order to facilitate the analysis of the effective capacity. Based on the proposed lower bound, performance of dense next generation heterogeneous cellular networks under  statistical QoS requirements was analyzed by building a scalable mathematical framework.

The authors in \cite{Wenchi} proposed a QoS-driven power allocation scheme for full-duplex wireless links with the goal of maximizing the overall effective capacity under a given delay QoS constraint. Two models namely local transmit power related self-interference (LTPRS) model and local transmit power unrelated self-interference (LTPUS) were built to analyze the full-duplex transmission, respectively. However, an approximation of the sum Shannon capacity was used in the formulation of the effective capacity under the assumption that the signal-to-interference-plus-noise ratio is much larger than 1. \cite{YuWang} considered the problem of distributed power allocation in a full duplex (FD) wireless network consisting of multiple pairs of nodes with the goal of maximizing the network-wide capacity. Shannon capacity was used as the performance metric and the optimal transmission powers for the FD transmitters were derived based on the high SINR approximation and a more general approximation method for the logarithm function.

The problem of joint subchannel allocation and power control was discussed in many studies. For instance, resource allocation in multicell uplink orthogonal frequency division multiple access (OFDMA) systems was considered in \cite{Buzzi}, and the problem was solved via noncooperative games for subcarrier allocation and transmit power control. \cite{Jun} proposed a joint power control and subchannel allocation for OFDMA femtocell networking using distributed auction game in order to minimize the total power radiated by the femtocell base station and guaranteeing the throughput. \cite{Thanabalasingham} considered the problem of joint subcarrier and power allocation for the downlink of a multiuser OFDM cellular network in order to minimize the power consumption subject to meeting the target rates of all users in the network. The authors in \cite{Di} considered the adaptive subcarrier assignment and fair power control strategy that minimize a cost function of average relay powers in multiuser wireless OFDM networks.

However, the aforementioned works have not considered statistical QoS requirements, bandwidth limitations, power limitations and interference jointly in FD wireless networks. In this paper, we address the problem of maximizing the weighted sum quality of reconstructed videos at the receivers subject to total bandwidth, minimum video quality, maximum transmission power and delay QoS constraints by allocating the bandwidth and determining the optimal power level for each user when statistical channel side information (CSI) is available in the FD wireless network. Since the optimization problem is neither a concave nor convex problem due to the existence of the interference, we employ the monotonic optimization (MO) framework. Our more specific contributions include the following:
\begin{enumerate}
\item We reformulate the optimization problem as a monotonic optimization problem, and propose a  framework to study full-duplex communication via monotonic optimization.

\item We derive several key properties of the optimal solution space.

\item We develop algorithms to efficiently determine the optimal resource allocation policies. In particular, we develop algorithms for enclosing polyblock initialization, projection onto the upper boundary, and iterative derivation of new enclosing polyblocks.

\item We analyze the impact of important system parameters (e.g., video quality parameters, QoS constraints, and weights) on the optimal resource allocation strategies and received video quality in terms of peak signal-to-noise ratio.

\end{enumerate}

The remainder of this paper is organized as follows: The system model is presented in Section \ref{sec:System_Model}.  Statistical QoS guarantees, effective capacity as a throughput metric, and quality-rate model are described as preliminary concepts in Section \ref{sec:Preliminary}. The optimization problems are formulated and the optimal policies are derived in Section \ref{sec:Maximization}. Simulation results are presented and discussed in Section \ref{sec:Result}. Finally, we conclude the paper in Section \ref{sec:Conclusion}. Proofs are relegated to the Appendix.

\section{System Model} \label{sec:System_Model}
\begin{figure*}
\centering
\includegraphics[width=0.8\textwidth]{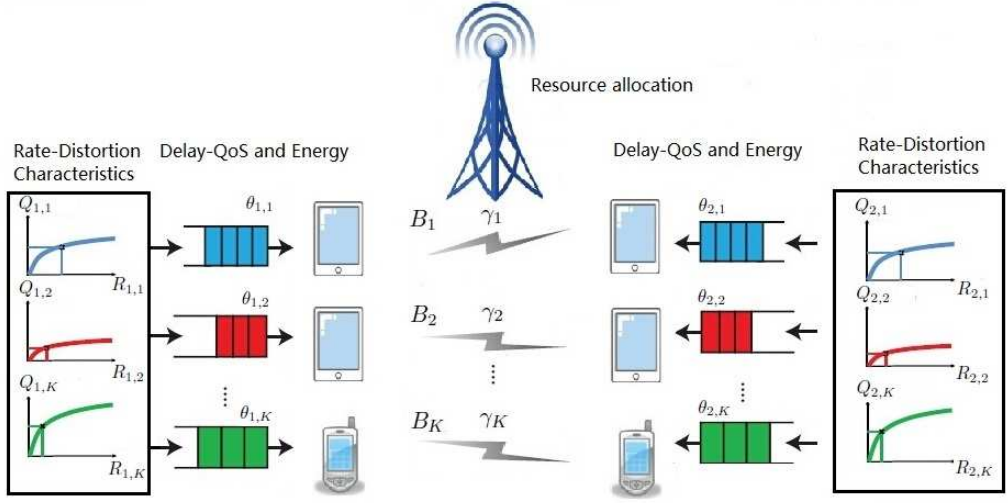}
\caption{Wireless system model in which each pair of users communicates in full-duplex mode under quality and delay constraints.}\label{fig:System_Model}
\end{figure*}

Fig. \ref{fig:System_Model} depicts the considered system model. We consider $K$ pairs
 of users, denoted as $(U_{1, 1}, U_{2, 1})$, $(U_{1, 2}, U_{2, 2})$, \ldots , $(U_{1, K}, U_{2, K})$\footnote{Throughout the paper, the subscripts $(1,k)$ and $(2,k)$ are used for parameters and notations related to users 1 and 2 of the $k^{th}$ pair, respectively.}, orthogonally sharing a total bandwidth of $B$ Hz in FD mode. Specifically, the $k$th FD link between $U_{1, k}$ and $U_{2, k}$ is allocated a bandwidth of $B_k$ Hz for the transmission of the video data under the constraint that the total bandwidth is $\sum_{k=1}^K B_k = B$. It is assumed that flat fading is experienced in each subchannel. The channel coherence time is denoted by $T_c$, and the timescale of video rate adaptation is much larger than $T_c$ in practice for video transmission since video source rate is adapted at the group of pictures (GOP) time scale which is measured in seconds. The case in which the channel state changes faster than the source rate is considered in our system since if the fading channel state varies at the same timescale as the source rate, statistical delay guarantees become less interesting \cite{Khalek}.

The practical application of this model includes, for instance, scenarios in which device-to-device (D2D) users exchange multimedia data (e.g., via social media sites) or conduct teleconferencing (i.e., engage in interactive video) in full-duplex mode. Assuming the availability of only statistical channel side information (CSI), base station acts as a coordinating agent and performs quality-driven resource allocation. Or in a different scenario, we can have one base station performing full-duplex multimedia communication with multiple users over different subchannels (e.g., via othogonal frequency division multiple access (OFDMA)). In this case, all the users on the left-hand side of Fig. \ref{fig:System_Model} essentially represent (or collapse to) a single base station in which there are multiple buffers and multiple flows of multimedia data to be sent to different users on the right-hand side. Base station again performs quality-driven resource allocation.

\section{Preliminaries} \label{sec:Preliminary}
\subsection{Notations}
Throughout this paper, vectors are denoted by boldface letters, the $j$-th entry of a vector $\mathbf{x}$ is denoted by $x_j$. $\mathcal{R}$ and $\mathcal{R}_+$ denote the set of real numbers and nonnegative real numbers, respectively. $\mathcal{R}^n$ and $\mathcal{R}_+^n$ denote the space of $n$-dimensional real-valued vectors and nonnegative real-valued vectors, respectively.
For any two vectors $\mathbf{x}$, $\mathbf{y} \in \mathcal{R}^n$, $\mathbf{x} \geq \mathbf{y}$ if $x_j \geq y_j$  for all $j = 1, 2, \ldots,n$. $\cup$, $\cap$ and $\setminus$ represent set union, set intersection and set difference operators, respectively. $\mathbf{e}_j \in \mathcal{R}^n$ denotes the $j$-th unit vector of $\mathcal{R}^n$, i.e., the vector such that $e_j = 1$ and $e_i = 0$ for all $i \neq j$.

\subsection{Delay QoS Constraints and Effective Capacity}
In wireless video transmissions, queue length in the buffer is subject to limitations to control the queueing delay. In particular, we assume that the overflow probabilities in the buffer storing the data to be transmitted at each pair of users decay exponentially for large buffer threshold, i.e.,
\begin{align}
& \Pr\{l_{i,k}>l_{i,k}^{th}\} \approx e^{-\theta_{i,k} l_{i,k}^{th}},  i \in \mathcal{I}, k \in \mathcal{K}
\end{align}
where $l_{i,k}$ and $l_{i,k}^{th}$ are the queue length and threshold at $U_{i,k}$, respectively, and $\mathcal{I} = \{1, 2\}$ and $\mathcal{K} = \{1, 2,\ldots,K\}$. $\theta_{i, k}$, referred to as the QoS exponent, determines the decay rate of the buffer overflow probability, and characterizes how strict the queueing/delay constraints are. Larger $\theta_{i,k}$ leads to more stringent QoS requirements while smaller $\theta_{i,k}$ represents looser QoS requirements.  In the presence of such QoS requirements, two key performance metrics are effective capacity and effective bandwidth. Effective capacity (EC), $C(\theta_{i,k})$, characterizes the maximum constant arrival rate which can be supported by the service process (i.e., wireless transmissions) in the presence of statistical buffer overflow constraints specified by the QoS exponent $\theta_{i,k}$. Effective bandwidth, $A(\theta_{i,k})$, provides the minimum constant service rate needed to guarantee that the overflow probability decades with rate specified by $\theta_{i,k}$ for the given arrival process $\{A\}$.

Now, we express the EC formulations for the pair of users operating in FD mode. Considering independent and identically distributed fading in each coherence block of duration $T_c$, we can write the EC expressions for the $k^{th}$ pair of users as
\begin{align}
C_{1,k}(\theta_{1,k}) & \nonumber = -\frac{1}{\theta_{1,k} T_c}\ln\left(\mathbb{E}_{\gamma_k}\{e^{-\theta_{1,k} r_{1,k}}\}\right) \\
& = -\frac{1}{\theta_{1,k} T_c}\ln\left(\!\!\mathbb{E}_{\gamma_k}\!\!\left\{e^{-\theta_{1,k} B_k T_c \log\left(1+\frac{P_{1,k} \gamma_k}{N_0 B_k + I_{2,k}}\right)}\right\}\right) \label{EC1}\\
C_{2,k}(\theta_{2,k}) & \nonumber = -\frac{1}{\theta_{2,k} T_c}\ln\left(\mathbb{E}_{\gamma_k}\{e^{-\theta_{2,k} r_{2,k}}\}\right) \\
& = -\frac{1}{\theta_{2,k} T_c}\ln\left(\!\!\mathbb{E}_{\gamma_k}\!\!\left\{e^{-\theta_{2,k} B_k T_c \log\left(1+\frac{P_{2,k} \gamma_k}{N_0 B_k + I_{1,k}}\right)}\right\}\right) \label{EC2}
\end{align}
where $B_k$ is the allocated bandwidth for the full duplex communication of these users, $P_{i,k}$ is the power of user $U_{i,k}$, and $\theta_{i,k}$ is the QoS exponent of $U_{i,k}$. Moreover, $N_0$ is the power spectral density of the background Gaussian noise, and $I_{1,k}$ and $I_{2,k}$ are the self-interference terms at $U_{1,k}$ and $U_{2,k}$, respectively.


The EC should be equal to the effective bandwidth of the arrival process for the given QoS exponent $\theta$ \cite{Qinghe} in order to support the highest arrival rates. For constant arrival rate $R$, the effective bandwidth of the arrival process is $A(\theta_{i,k}) = R$. Therefore, the maximum constant arrival rates at users $U_{1,k}$ and $U_{2,k}$ can be expressed, respectively, as
\begin{align}
R_{1,k} & \nonumber = A_{1,k}(\theta_{1,k}) = C_{1,k}(\theta_{1,k}) \\
    & = -\frac{1}{\theta_{1,k} T_c}\ln\left(\!\mathbb{E}_{\gamma_k}\!\!\left\{e^{-\theta_{1,k} B_k T_c \log\left(1+\frac{P_{1,k} \gamma_k}{N_0 B_k + I_{2,k}}\right)}\right\}\right), \label{R_1k1} \\
R_{2,k} & \nonumber = A_{2,k}(\theta_{2,k}) = C_{2,k}(\theta_{2,k}) \\
    &-\frac{1}{\theta_{2,k} T_c}\ln\left(\!\mathbb{E}_{\gamma_k}\!\!\left\{e^{-\theta_{2,k} B_k T_c \log\left(1+\frac{P_{2,k} \gamma_k}{N_0 B_k + I_{1,k}}\right)}\right\}\right). \label{R_2k1}
\end{align}

\subsection{Video Quality-Rate Model}
Lossy data compression, which focuses on the tradeoff between the distortion and bit rate, is used in video coding algorithms, where an increased distortion leads to a decreased rate and vice-versa. Rate-distortion (R-D) theory addresses the problem of determining the minimal source bit rate so that the distortion of the reconstructed data at the receiver does not exceed a given distortion value. Thus, the R-D function can estimate the bit rate at given distortion, or estimate the distortion at a given bit rate. Moreover, operational R-D (ORD) theory is applied to lossy data compression with finite number of possible R-D pairs, and the ORD function shows that the bit rate is a convex function of distortion.
In \cite{Chao}, the quality of video is measured in terms of the reversed difference mean opinion score (RDMOS), and the following rate-quality model to predict $q_u(t)$ using the video data rate $r_u(t)$ is employed:
\begin{equation}
q_u(t) = \alpha_u(t)\log(r_u(t)) + \beta_u(t)
\end{equation}
where model parameters $\alpha_u(t)$ and $\beta_u(t)$ can be determined by minimizing the prediction error.
Also several R-D models are proposed in \cite{Luis}, in which the quality is measured in terms of peak signal-to-noise ratio (PSNR). The exponential model for the rate-PSNR curve is used in our paper. Thus, PSNR-rate curve is described by a logarithmic model and can be expressed as follows:
\begin{align}
& Q_{i,k} = a_{i,k}\ln(R_{i,k}) + b_{i,k} \label{Qmodel1}
\end{align}
where $R_{i,k}$ and $Q_{i,k}$ are the arrival rate and PSNR of the transmitted video at $U_{i,k}$, respectively, and $a_{i,k}$ and $b_{i,k}$ are the parameters that can be determined by minimizing the prediction error. As discussed in the previous subsection, the source rate of the channel is given by the effective capacity (which quantifies the maximum constant arrival rate), i.e., $R_{i,k} = C_{i,k}$ in order to achieve the maximum video quality.

\section{Weighted Sum Quality-Maximizing Policies} \label{sec:Maximization}

In this section, optimization problems are formulated to maximize the weighted sum video quality subject to maximum transmission power and minimum video quality constraints at each user and a total bandwidth constraint. More specifically, we address the optimal allocation of bandwidth and the determination of transmission power levels assuming the availability of statistical CSI.
It is assumed that each user just has one antenna for transmitting and receiving the data. Thus, the self-interference just depends on the self-transmission power, and the maximum constant arrival rate in (\ref{R_1k1}) and (\ref{R_2k1}) can be rewritten as
\begin{align}
R_{1,k} & \nonumber = C_{1,k}(\theta_{1,k}) \\
    & = -\frac{1}{\theta_{1,k} T_c}\ln\left(\mathbb{E}_{\gamma_{k}}\left\{ e^{-\theta_{1,k} B_k T_c \log\left(1+\frac{P_{1,k} \gamma_{k}}{N_0 B_{k} + \mu_{2,k}P_{2,k}}\right)}\right\}\right) \label{R_1k2}\\
R_{2,k} & \nonumber = C_{2,k}(\theta_{2,k}) \\
    & = -\frac{1}{\theta_{2,k} T_c}\ln\left(\mathbb{E}_{\gamma_{k}}\left\{ e^{-\theta_{2,k} B_k T_c \log\left(1+\frac{P_{2,k} \gamma_{k}}{N_0 B_{k} + \mu_{1,k}P_{1,k}}\right)}\right\}\right) \label{R_2k2}
\end{align}
where $\mu_{i,k}\in (0, 1]$ is the self-interference suppression factor at $U_{i,k}$. We can now express the weighted sum video quality at users $U_{1,k}$ and $U_{2,k}$ as
\begin{align}
\nonumber Q_{k} = & \omega_{1,k}Q_{1,k} + \omega_{2,k}Q_{2,k} \\
= & \sum_{i=1}^2 \omega_{i,k}\big(a_{i,k}\ln(R_{i,k}) + b_{i,k}\big), \label{Q_sk}
\end{align}
where $\omega_{i,k} \in [0, 1]$ denotes the weight for the quality of the video transmitted by user $U_{i,k}$ such that $\sum_{k=1}^K \sum_{i=1}^2 \omega_{i,k} = 1$.

Now, the problem of maximizing the overall sum video quality of all users over bandwidth and power allocation strategies can be expressed as follows:
\begin{subequations}
\begin{align}
\max_{\substack{\mathbf{B}, \mathbf{P_{1}}, \mathbf{P_{2}}}} & \sum_{k=1}^{K}\sum_{i=1}^{2}\Big(\omega_{i,k} Q_{i,k}(R_{i,k}) \Big) \label{Prob_Const1}\\
\text{s.t.} \quad & \sum_{k=1}^{K}B_k \leq B; \quad  B_k \geq 0, \quad \forall k \in \mathcal{K} \label{Band_Const1}\\
& P_{i,k} \leq P_{i,k}^{max}; \quad  P_{i,k} \geq 0, \quad \forall i \in \mathcal{I},k \in \mathcal{K} \label{Power_Const1} \\
& Q_{i,k}(R_{i,k}) \geq Q_{i,k}^{min}, \quad \forall i \in \mathcal{I},k \in \mathcal{K} \label{Quality_Const1}
\end{align}
\label{Prob_Const1}
\end{subequations}

Above, (\ref{Band_Const1}) is the total bandwidth constraint, (\ref{Power_Const1}) is the maximum transmission power constraint at each user and (\ref{Quality_Const1}) is the minimum required video quality constraint. Specifically, $P_{i,k}^{max}$ and $Q_{i,k}^{min}$ are the maximum available transmission power and minimum transmitted video quality at $U_{i,k}$, respectively. $\mathbf{B}$, $\mathbf{P}_1$ and $\mathbf{P}_2$ are $K\times 1$ vectors of bandwidth allocated to each link, power allocated to $U_{1, k}$ and $U_{2, k}$, respectively. The feasible set of $\mathbf{B}$ is denoted by $\mathcal{B} = \{\mathbf{B}|\sum_{k=1}^{K}B_k \leq B\}$, and the feasible sets of $\mathbf{P}_1$ and $\mathbf{P}_2$ are denoted by $\mathcal{P}_1 = \{\mathbf{P}_1|P_{1,k} \leq P_{1,k}^{max}, \forall k \in \mathcal{K}\}$ and $\mathcal{P}_2 = \{\mathbf{P}_2|P_{2,k} \leq P_{2,k}^{max}, \forall k \in \mathcal{K}\}$, respectively.

\subsection{Problem Reformulation as Monotonic Optimization}

First, we introduce some definitions used in monotonic optimization (MO) from \cite{Zhang}, and then show that problem (\ref{Prob_Const1}) can be reformulated as an MO problem.
\begin{definition}
\emph{(Box)}
\label{box}
For two vectors $\mathbf{a}\in\mathcal{R}^n$, $\mathbf{b}\in\mathcal{R}^n$ with $\mathbf{a}\leq\mathbf{b}$, the box $[\mathbf{a}, \mathbf{b}]$ is the set of all vectors $\mathbf{x}\in\mathcal{R}^n$ satisfying $\mathbf{a}\leq\mathbf{x}\leq\mathbf{b}$. In other words, a hyperrectangle $[\mathbf{a}, \mathbf{b}]= \{\mathbf{x}|a_j \leq x_j \leq b_j, j = 1, 2, ..., n\}$ is referred as a box.
\end{definition}

\begin{definition}
\emph{(Normal set)}
\label{Nset}
A set $\mathcal{G} \subset \mathcal{R}_+^n$ (the $n$-dimensional nonnegative real domain) is normal if for any element $\mathbf{x}\in\mathcal{G}$, all other elements $\mathbf{x^\prime}$ such that $\mathbf{0} \leq \mathbf{x^\prime} \leq \mathbf{x}$ are in the same set $\mathcal{G}$. In other words, $\mathcal{G} \subset \mathcal{R}_+^n$ is normal if for any $\mathbf{x}\in\mathcal{G}$, the set $[\mathbf{0}, \mathbf{x}] \subset \mathcal{G}$.
\end{definition}

\begin{definition}
\emph{(Conormal set)}
\label{CNset}
A set $\mathcal{H} \subset \mathcal{R}_+^n$ is conormal if for any element $\mathbf{x}\in\mathcal{H}$, all other elements $\mathbf{x^\prime}$ such that $\mathbf{x^\prime} \geq \mathbf{x}$ are in the same set $\mathcal{H}$. In other words, a set $\mathcal{H}$ is conormal in $[\mathbf{0}, \mathbf{b}]$ if for any $\mathbf{x}\in\mathcal{H}$, $[\mathbf{x}, \mathbf{b}] \subset \mathcal{H}$.
\end{definition}

\begin{definition}
\emph{(Upper boundary)}
\label{UPbound}
An element $\mathbf{\bar{x}}$ of a normal closed set $\mathcal{G}$ is an upper boundary point of $\mathcal{G}$ if $\mathcal{G} \cap \{\mathbf{x} \in \mathcal{R}_+^n|\mathbf{x} > \mathbf{\bar{x}}\} = \emptyset$. The set of all upper boundary points of the set $\mathcal{G}$ is called its upper boundary and denoted by $\partial^+\mathcal{G}$.
\end{definition}

\begin{definition}
\emph{(Polyblocks)}
\label{Pblock}
A set $\mathcal{S} \subset \mathcal{R}_+^n$ is a polyblock if it is a union of a finite number of boxes $[\mathbf{0}, \mathbf{z}]$, where $\mathbf{z} \in \mathcal{T}$ and $|\mathcal{T}| < +\infty$. The set $\mathcal{T}$ is the vertex set of the polyblock.
\end{definition}

\begin{definition}
\emph{(Proper)}
\label{Proper}
An element $\mathbf{x} \in \mathcal{T}$ is said to be proper if there is no $\mathbf{x}^\prime \in \mathcal{T}$ such that $\mathbf{x}^\prime \neq \mathbf{x}$ and $\mathbf{x}^\prime \geq \mathbf{x}$. If every element $\mathbf{x}^\prime \in \mathcal{T}$ is proper, then the set $\mathcal{T}$ is a proper set.
\end{definition}

From \cite{Zhang}, an optimization problem belongs to the class of MO if it can be represented in the following form:
\begin{align}
& \max{f(\mathbf{x})} \label{Mono_1}   \\
\text{s.t.} & \quad \mathbf{x} \in \mathcal{G} \cap \mathcal{H}
\end{align}
where $f(\mathbf{x}):\mathcal{R}_+^n \rightarrow \mathcal{R}$ is an increasing function, $\mathcal{G} \subset [\mathbf{0}, \mathbf{b}] \subset \mathcal{R}_+^n$ is a compact normal set, and $\mathcal{H}$ is a closed conormal set on $[\mathbf{0}, \mathbf{b}]$. A simpler case is the one in which $\mathcal{H}$ is not present in the formulation (which occurs e.g., if the conormal set $\mathcal{H}$ is box $[\mathbf{0}, \mathbf{b}]$). In general, if $\mathcal{G} \cap \mathcal{H} \neq \emptyset$, the problem is considered feasible.

We note that it is not possible to obtain the optimal solution of (\ref{Prob_Const1}) based on the theory of convex optimization \cite{Boyd} because of the non-convexity of the optimization problem in (\ref{Prob_Const1}) in terms of $P_{i,k}$ and $B_k$ jointly. This non-convexity is primarily due to the presence of the self-interference terms. In operations research, monotonicity is regarded as another important property for effectively solving an optimization problem. Therefore, we follow the approach to solve the non-convex problem (\ref{Prob_Const1}) by transforming it into an MO problem, and then solving the corresponding MO problem based on recent advances in monotonic optimization \cite{Tuy}.

Let $\mathbf{Y}$ denote the vector $(Y_1, Y_2, \ldots, Y_{2K})$ with $Y_j$ being the $j$-th component of $\mathbf{Y}$. We define the function
\begin{gather*}
\Phi(\mathbf{Y}) = \sum_{k=1}^{K}\sum_{i=1}^{2} \omega_{i,k}\left[ a_{i,k}\ln\left(\frac{1}{\theta_{i,k}T_c}\ln(Y_{(i-1)K+k})\right) + b_{i,k}\right].
\end{gather*}
It is easy to see that $\Phi(\mathbf{Y})$ is an increasing function of $\mathbf{Y}$ on $\mathcal{R}_+^{2\times K}$.
In other words, for any two vectors $\mathbf{Y}_1$ and $\mathbf{Y}_2$, $\Phi(\mathbf{Y}_1) \geq \Phi(\mathbf{Y}_2)$ if $\mathbf{Y}_1 \geq \mathbf{Y}_2$. Now, problem (\ref{Prob_Const1}) can be rewritten in the MO formulation as
\begin{subequations}
\begin{align}
\max \Phi(\mathbf{Y}) = &\sum_{k=1}^{K}\sum_{i=1}^{2} \omega_{i,k}\left[a_{i,k}\ln\left(\frac{\ln(Y_{(i-1)K+k})}{\theta_{i,k}T_c}\right) + b_{i,k}\right] \label{MO_fun1} \\
\text{s.t.} \quad & \mathbf{Y} \in \mathcal{G} \cap \mathcal{H}.\label{eq:normal-conormal-sets}
\end{align}
\label{MO_fun1}
\end{subequations}
Above, the normal set is
\begin{align}
\mathcal{G} = \left\{\mathbf{Y}|0 \leq Y_{(i-1)K+k} \leq V_{(i-1)K+k}(P_{1,k}, P_{2,k}, B_k), \forall i \in \mathcal{I}, \right.\nonumber \\ \left.\forall k \in \mathcal{K}, \mathbf{P}_1 \in \mathcal{P}_1, \mathbf{P}_2 \in \mathcal{P}_2, \mathbf{B} \in \mathcal{B}\right\}
\end{align}
where
\begin{align}
\nonumber  V_{(i-1)K+k}&(P_{1,k},P_{2,k},B_k)   \\= & \left(\mathbb{E}_{\gamma_{k}}\left\{ e^{-\theta_{i,k} B_k T_c \log\left(1+\frac{P_{i,k} \gamma_{k}}{N_0 B_{k} + \mu_{3-i,k}P_{3-i,k}}\right)}\right\}\right)^{-1}. \label{V}
\end{align}
Note that when $ Y_{(i-1)K+k}$ in the objective function in (\ref{MO_fun1}) is replaced with the upper bound $V_{(i-1)K+k}(P_{1,k},P_{2,k},B_k)$, the objective function becomes the same as that in (\ref{Prob_Const1}).
In (\ref{eq:normal-conormal-sets}), the conormal set is
\begin{align}
\mathcal{H} = \{\mathbf{Y}| Y_{(i-1)K+k} \geq V_{(i-1)K+k}^{min}, \forall i \in \mathcal{I}, \forall k \in \mathcal{K} \}
\end{align}
where $$V_{(i-1)K+k}^{min} = e^{\theta_{i,k}T_c e^{\frac{Q_{i, k}^{min}-b_{i,k}}{a_{i,k}}}}.$$
Note that the normal set $\mathcal{G}$ describes the combination of total bandwidth constraint (\ref{Band_Const1}) and maximum transmission power constraint (\ref{Power_Const1}), and the cornormal set $\mathcal{H}$ corresponds to the minimum quality constraint (\ref{Quality_Const1}).

Since $\Phi(\mathbf{Y})$ is an increasing function of $\mathbf{Y}$, the optimal solution of Problem (\ref{MO_fun1}), denoted by $\mathbf{Y}^*$, must be  located at the upper boundary of $\mathcal{G}$, denoted by $\partial^+\mathcal{G}$. This means that we can find a bandwidth allocation $\mathbf{B}^*$ and power allocations $\mathbf{P}_1^*$ and $\mathbf{P}_2^*$ corresponding to the optimal solution $\mathbf{Y}^*$ such that
\begin{gather}
Y_{(i-1)K+k}^* = \left(\mathbb{E}_{\gamma_{k}}\left\{ e^{-\theta_{i,k} B_k^* T_c \log\left(1+\frac{P_{i,k}^* \gamma_{k}}{N_0 B_{k}^* + \mu_{3-i,k}P_{3-i,k}^*}\right)}\right\}\right)^{-1}
\end{gather}
for all $i \in \mathcal{I}$ and $k \in \mathcal{K}$. Therefore, such $\mathbf{B}^*$, $\mathbf{P}_1^*$ and $\mathbf{P}_2^*$ are clearly the optimal solutions to Problem (\ref{Prob_Const1}). Hence Problem (\ref{Prob_Const1}) and (\ref{MO_fun1}) are equivalent.
We must also note that $Y_{(i-1)K+k}$ is lower bounded by 1, i.e., $Y_{(i-1)K+k} \geq 1$ for all $i$ and $k$. Consequently, the optimal solution $\mathbf{Y}^*$ to Problem (\ref{MO_fun1}), which is located only at the upper boundary of set $\mathcal{G}$, is also lower bounded by $\mathbf{1}$. That means that the optimal solution $\mathbf{Y}^* \in \mathcal{G} \cap \mathcal{H} \cap \mathcal{L}$, where $$\mathcal{L} = \{\mathbf{Y}|Y_{(i-1)K+k}\geq 1,\forall i \in \mathcal{I}, \forall k \in \mathcal{K}\}.$$

\subsection{Initialization of the Enclosing Polyblock}
In order to better approximate the upper boundary of the feasible set, we need to initialize the polyblock that contains the feasible set properly. In other words, we need to find the smallest box $[\mathbf{0}, \mathbf{v}^\prime]$ that contains $\mathcal{G} \cap \mathcal{H} \cap \mathcal{L}$. Since both sets $\mathcal{H}$ and $\mathcal{L}$ are cornormal, the set $$\mathcal{J} = \mathcal{H} \cap \mathcal{L} = \{\mathbf{Y}| Y_{(i-1)K+k} \geq \max\{V_{(i-1)K+k}^{min}, 1\}, \forall i \in \mathcal{I}, \forall k \in \mathcal{K} \}$$ is also cornormal. The smallest $\mathbf{v}^\prime$ such that $[\mathbf{0}, \mathbf{v}^\prime]$ contains $\mathcal{G} \cap \mathcal{J}$ is given by the following:
\begin{align}
v_j^\prime = \max \{Y_j|\mathbf{Y} \in \mathcal{G} \cap \mathcal{J}\} \quad \forall j = 1,\ldots,2K .
\end{align}
Before describing the enclosing polyblock initialization algorithm, we provide the following characterization for the functional properties of $V_{(i-1)K+k}(P_{1,k},P_{2,k},B_k)$.
\begin{Theo} \label{theo:min_bandw}
Consider the functions
\begin{align}
V_1(P_1, P_2, B) &= \left(\mathbb{E}_{\gamma}\left\{ e^{-\theta B T_c \log\left(1+\frac{P_1 \gamma}{N_0 B + \mu P_2}\right)}\right\}\right)^{-1} \text{and}
\\
V_2(P_1, P_2, B) &= \left(\mathbb{E}_{\gamma}\left\{ e^{-\theta B T_c \log\left(1+\frac{P_2 \gamma}{N_0 B + \mu P_1}\right)}\right\}\right)^{-1}
\end{align}
and assume that $P_1 \le P^{\max}$ and $P_2 \le P^{\max}$. Then, we have the following properties:
\begin{enumerate}
\item For given bandwidth $B$, $V_1$ is maximized when either $P_1 = P^{\max}$ or $P_2 = P^{\max}$. Hence, at least one power value should be at its maximum level.
\item For given $P_1$ and $P_2$, $V_1$ is an increasing function of $B$.
\item The above properties hold for $V_2$ as well due to the similarity in their definitions (with only roles of $P_1$ and $P_2$ switched).
\item The bandwidth required to achieve two target values $V_1(P_1, P_2, B) = V_1^*$ and $V_2(P_1, P_2, B) = V_2^*$ is minimized if either $P_1 = P^{\max}$ or $P_2 = P^{\max}$.
\end{enumerate}
\end{Theo}
\emph{Proof:} See Appendix \ref{Proof:1}.

The detailed algorithm for initializing the enclosing polyblock is provided below in Algorithm \ref{algorithm1}. We note that Step 3 of Algorithm 1 makes use of Theorem \ref{theo:min_bandw}, i.e., the fact that the minimum bandwidth always occurs at $P_{1,k} = P_{1,k}^{max}$ or $P_{2,k} = P_{2,k}^{max}$.
\begin{algorithm}[H]
\caption{The enclosing polyblock initialization algorithm}
\begin{algorithmic}[1]\label{algorithm1}
\REQUIRE $\mathcal{G}$, $\mathcal{H}$ and $\mathcal{L}$
\ENSURE Polyblock $\mathcal{S}_1$
\STATE Initialize $s = 1$.
\FOR {$k = 1:K$}
    \STATE Set $V_{(i-1)K+k}(P_{1,k}, P_{2,k}, B_k) = \max \{V_{(i-1)K+k}^{min}, 1\}$ for $i = 1, 2$. Let $P_{1,k} = P_{1,k}^{max}$, find the bandwidth $B_k = B_{k1}$ and power $P_{2,k}$ by solving (\ref{V}). Similarly, let $P_{2,k} = P_{2,k}^{max}$, find the bandwidth $B_k = B_{k2}$ and $P_{1,k}$ by solving (\ref{V}). $B_k^{min} = \min \{B_{k1}, B_{k2}\}$ if both $P_{1,k} \in \mathcal{P}_1$ and $P_{2,k} \in \mathcal{P}_2$, and $B_k^{min} = B_{ki}$ if just one $P_{i,k} \in \mathcal{P}_i$ for $i = 1$ or $i = 2$. Otherwise, Problem (\ref{MO_fun1}) does not have solution and set $s = 0$.
\ENDFOR
\STATE If $s = 1$, and $\sum_{l=1}^K B_l^{min} > B$, the Problem (\ref{MO_fun1}) does not have solution and set $s = 0$.
\IF {$s = 1$}
    \FOR {$k = 1:K$}
        \STATE $B_k = B - \sum_{l\neq k} B_l^{min}$.
        \FOR {i = 1:2}
             \STATE Let $V_{(2-i)K+k} = \max \{V_{(2-i)K+k}^{min}, 1\}$ and $P_{3-i,k} = P_{3-i,k}^{max}$, find the power $P_{i,k}$ by solving (\ref{V}).
             \STATE Calculate $V_{(i-1)K+k}^{max}$ from (\ref{V}) by substituting $P_{j,k}$ and $P_{i,k}$ obtained above.
        \ENDFOR
    \ENDFOR
\ENDIF
\STATE Therefore, the vector $\mathbf{v^\prime} = (V_{1}^{max},\ldots, V_{2K}^{max})$ is the vertex of the initial polyblock $\mathcal{S}_1$.
\end{algorithmic}
\end{algorithm}

\begin{figure}
\centering
\includegraphics[width=0.3\textwidth]{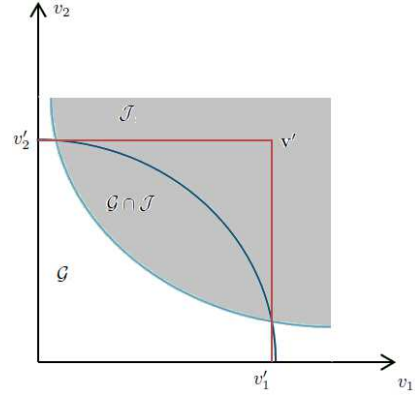}
\caption{Example of initialized enclosing polyblock}\label{fig:Polyblock}
\end{figure}

We now provide an illustration for the enclosing polyblock initialization. For instance, assume that $\mathcal{G}$ and $\mathcal{J}$ are two-dimensional sets by assuming $K = 1$. As shown in Fig. \ref{fig:Polyblock}, the box $[\mathbf{0}, \mathbf{v}^\prime]$ constrained by the red lines is the smallest box that contains $\mathcal{G} \cap \mathcal{J}$, where $\mathbf{v}^\prime = (v_1^\prime, v_2^\prime)$. And $\mathbf{v}^\prime$ can be obtained by the algorithm provided above.

Before we solve the optimization problem by using MO theory, we provide the following proposition from \cite{Zhang}.
\begin{Prop}
\label{Prop1}
\emph{(Projection on the upper boundary) [17]}
Let $\mathcal{G} \subset \mathcal{R}_+^n$ be a compact normal set with nonempty interior. Then, for any point $\mathbf{x} \in \mathcal{R}_+^n \setminus \mathcal{G}$, the line connecting $\mathbf{0}$ and $\mathbf{x}$ intersects the upper boundary $\partial^+\mathcal{G}$ of $\mathcal{G}$ at a unique point $\pi_{\mathcal{G}}(\mathbf{x})$, which is defined as
\begin{align}
\pi_{\mathcal{G}}(\mathbf{x})= \lambda \mathbf{x}, \, \text{ where } \lambda = \arg\max \{\alpha>0 \mid \alpha\mathbf{x} \in \mathcal{G}\}.
\end{align}
$\pi_{\mathcal{G}}(\mathbf{x})$ is the projection of $\mathbf{x}$ on the upper boundary $\partial^+\mathcal{G}$.
\end{Prop}

Due to the presence of $\mathcal{J}$, $\pi_{\mathcal{G}(\mathbf{x})}$ may be located outside the feasible set $\mathcal{G} \cap \mathcal{J}$ if one end point of the line is $\mathbf{0}$. In order to avoid this situation, we modify the projection by changing the line connecting $\mathbf{0}$ and $\mathbf{x}$ to the line connecting $\mathbf{u}$ and $\mathbf{x}$, and we denote by $\pi_{\mathcal{G}}^{\mathbf{u}}(\mathbf{x})$ the projection of $\mathbf{x}$ on the upper boundary $\partial^+\mathcal{G}$ with $\mathbf{u}$ acting as the origin. Therefore,
\begin{align}
\pi_{\mathcal{G}}^{\mathbf{u}}(\mathbf{x}) = \lambda (\mathbf{x}-\mathbf{u}) + \mathbf{u},
\end{align}
where $\lambda = \arg\max \{\alpha>0 \mid \alpha\mathbf{x} \in \mathcal{G}\}$ and $\mathbf{u} = (\max\{V_{1}^{min}, 1\}, \ldots, \max\{V_{2K}^{min}, 1\})$.

\subsection{Algorithms and Optimal Solution via Monotonic Optimization}
After obtaining the proper initial polyblock, we next develop algorithms and determine the optimal solution to Problem (\ref{MO_fun1}) via MO approach. The key idea of MO is to iteratively derive a new enclosing polyblock $\mathcal{S}_{j+1}$ from the previous polyblock $\mathcal{S}_{j}$ by cutting off the points that is in the infeasible set until reaching the $\epsilon$-error-tolerance solution. Following Proposition 3.8 in \cite{Zhang}, we let $\mathcal{S} \subset \mathcal{R}_+^n$ be a polyblock with a proper vertex set $\mathcal{T} \subset \mathcal{R}_+^n$ and let $\mathbf{x} \in \mathcal{S}$. Then, the new polyblock $\mathcal{S}_*$ has a vertex set
\begin{align}
\mathcal{T}^\prime =(\mathcal{T} \setminus \mathcal{T}_*)\cup\{\mathbf{v} = \mathbf{v} + (x_j - v_j)e_j|\mathbf{v} \in \mathcal{T}_*, j\in\{1,\ldots,n\}\}\label{Cutting}
\end{align}
where $\mathcal{T}_*$ is the subset of $\mathcal{T}$, consisting of the vertices at which $\Phi(\mathbf{Y})$ is maximized.
It is easy to see that if $\mathcal{S}$ is the proper polyblock such that $\mathcal{G} \cap \mathcal{J} \subset \mathcal{S}$ and $\mathbf{x} \in \partial^+\mathcal{G}$, then we have $\mathcal{G} \cap \mathcal{J} \subset \mathcal{S}_* \subset \mathcal{S}$.

We first construct a proper polyblock $\mathcal{S}_1$ that contains the feasible set, $\mathcal{G} \cap \mathcal{J}$ of Problem (\ref{MO_fun1}) by using Algorithm \ref{algorithm1}, and let $\mathcal{T}_1$ denote the initial proper vertex set of $\mathcal{S}_1$. There is just one vertex, $\mathbf{v}^\prime$, in $\mathcal{T}_1$. Since the objective function of Problem (\ref{MO_fun1}), $\Phi(\mathbf{Y})$ is monotonically increasing over set $\mathcal{S}_1$, the maximum of $\Phi(\mathbf{Y})$ occurs at some proper vertex $\mathbf{Y}_1$ of $\mathcal{S}_1$, i.e., $\mathbf{Y}_1 \in \mathcal{T}_1$. If $\mathbf{Y}_1$ is also in the feasible set $\mathcal{G} \cap \mathcal{J}$, then the optimization problem is solved and $\mathbf{Y}^* = \mathbf{Y}_1$. Otherwise, a smaller polyblock $\mathcal{S}_2 \subset \mathcal{S}_1$ is constructed such that $\mathcal{G} \cap \mathcal{J} \subset \mathcal{S}_2$ but excludes $\mathbf{Y}_1$ by using Proposition 3.8 in \cite{Zhang}. Therefore, a new vertex set $\mathcal{T}_2$ is constructed by replacing $\mathbf{Y}_1$ in $\mathcal{T}_1$ with $2\times K$ new vertices and removing the improper vertices. This procedure is repeated until an $\epsilon$-error-tolerance solution is found.
If $\mathbf{Y}_j$ denotes the optimal vertex that maximizes $\Phi(\mathbf{Y})$ over set $\mathcal{S}_j$ at the $j$-th iteration, we have $\mathcal{S}_1 \supset \mathcal{S}_2 \supset \cdots \supset \mathcal{G}$ and $\Phi(\mathbf{Y}_1) \geq \Phi(\mathbf{Y}_2) \geq \cdots \geq \Phi(\mathbf{Y}^*)$. $\mathbf{Y}_j^\prime =  \arg\max \{\Phi(\mathbf{Y})|\mathbf{Y} \in \{ \pi_{\mathcal{G}}^{\mathbf{u}}(\mathbf{Y}_j), \mathbf{Y}_{j-1}^\prime \} \}$ denotes the current best solution (CBS), and the current best value (CBV) is $\Phi(\mathbf{Y}_j^\prime)$ in the $j$-th iteration. Consequently, we have $\Phi(\mathbf{Y}_1^\prime) \leq \Phi(\mathbf{Y}_2^\prime) \leq \cdots \leq \Phi(\mathbf{Y}^*)$. The algorithm terminates at the $j$-th iteration if $\mathbf{Y}_j \in \mathcal{S}_j$, and $(1+\epsilon)\Phi(\mathbf{Y}_j^\prime) \geq \Phi(\mathbf{Y}_j)$ or $|\Phi(\mathbf{Y}_j^\prime) - \Phi(\mathbf{Y}_j)| \leq \epsilon$ based on the chosen strategy, where $\epsilon > 0$ is a small positive number representing the error tolerance. $\mathbf{Y}_j^\prime$ is the optimal $\epsilon$-error-tolerance solution.

\begin{algorithm}[H]
\caption{Projection algorithm (for finding $\pi_{\mathcal{G}}(\mathbf{Y}_j)$)}
\begin{algorithmic}[1]\label{algorithm2}
\REQUIRE $\mathbf{Y}_j$, $\mathcal{G}$
\ENSURE $\lambda_j$ such that $\lambda_j = \arg\max \{\lambda_j >0|\lambda_j \mathbf{Y}_j \in \mathcal{G}\}$
\STATE Initialize $\lambda_j = 0$
\FOR {$d = 0:2^K-1$}
    \STATE Let $c$ be a $K$-digit binary integer corresponding to $d$, and $c_l$ denote the $l$-th binary digit of $c$.
    \FOR {$k = 1:K$}
        \IF {$c_k = 0$}
            \STATE $P_{1,k} = P_{1,k}^{max}$
        \ELSE
            \STATE $P_{2,k} = P_{2,k}^{max}$
        \ENDIF
        \STATE From (\ref{V}), we set $V_{(i-1)K+k}(P_{1,k}, P_{2,k}, B_k) = \lambda_{j, d+1} (Y_{(i-1)K+k}^j-u_{(i-1)K+k})+u_{(i-1)K+k}$.
    \ENDFOR
    \STATE Set $\sum_{k=1}^{K}B_k = B$.
    \STATE Therefore, we get $2K+1$ equations, $K$ unknown power variables $P_{1,k}$ or $P_{2,k}$, $K$ unknown bandwidth variables $B_k$ for all $k = 1,\ldots,K$, and unknown variable $\lambda_{j,d+1}$. We can get the value of $\lambda_{j,d+1}$ by solving this $2K+1$ equations. If $P_{i,k} \leq P_{i,k}^{max}$ for all $i = 1, 2$ and $k = 1,\ldots,K$, $\lambda_j = \max\{\lambda_j, \lambda\}$.
\ENDFOR
\STATE $\pi_{\mathcal{G}}^{\mathbf{u}}(\mathbf{Y}_j) = \lambda_j (\mathbf{Y}_j-\mathbf{u})+\mathbf{u}$.
\end{algorithmic}
\end{algorithm}

As discussed in the previous subsection with Proposition \ref{Prop1}, iterations in finding $\{\mathbf{Y}_j\}$ involve projection on the upper boundary.  We provide our projection algorithm for finding $\pi_{\mathcal{G}}^{\mathbf{u}}(\mathbf{Y}_j)$  as Algorithm 2 above. In steps 6 and 8 of this algorithm, the reason for considering $P_{1, k}$ or $P_{2, k}$ to be at the maximum level for all $k = 1,\ldots,K$ and $\sum_{k=1}^{K}B_k = B$ is that $\pi_{\mathcal{G}}^{\mathbf{u}}(\mathbf{Y}_j)$ is attained at the upper boundary of $\mathcal{G}$, and the upper boundary $\partial^+\mathcal{G}$ is reached only if one of the users transmits at the peak power level. The proof for this characterization is provided in Appendix \ref{Proof:2}, which primarily follows from the results of Theorem \ref{theo:min_bandw}.

After having obtained the initial enclosing polyblock $\mathcal{S}_1$ and identified the algorithm for projection on the boundary, we can now iteratively derive a new enclosing polyblock $\mathcal{S}_{j+1}$ from
the previous polyblock $\mathcal{S}_{j}$ by using Algorithm \ref{algorithm3} below. Eventually, we obtain the $\epsilon$-error-tolerance solution after terminating the iteration under a certain condition.

\begin{algorithm}[H]
\caption{The optimal resource allocation algorithm}
\begin{algorithmic}[1]\label{algorithm3}
\REQUIRE Function $\Phi(\mathbf{Y}):\mathcal{R}_+^{2\times K}\rightarrow \mathcal{R}$, compact normal set $\mathcal{G} \subset \mathcal{R}_+^{2\times K}$, and a closed conormal set $\mathcal{J} \subset \mathcal{R}_+^{2\times K}$ such that $\mathcal{G} \cap \mathcal{J} \neq \emptyset$
\ENSURE An $\epsilon$ error tolerance solution $\mathbf{Y}^*$ and the corresponding $\mathbf{P}_1^*$, $\mathbf{P}_2^*$ and $\mathbf{B}^*$.
\STATE Initialization: Let the initial polyblock $\mathcal{S}_1$ be the box $[\mathbf{0}, \mathbf{b}]$ that encloses $\mathcal{G} \cap \mathcal{J}$ (This can be obtained by using Algorithm \ref{algorithm1}). The vertex set $\mathcal{T}_1 = {\mathbf{b}}$. $\epsilon > 0$ is a small positive number. CBV $\Omega_0 = 0$ and $j = 0$.
\REPEAT
    \STATE $j = j+1$.
    \STATE Select $\mathbf{Y}_j \in \arg\max\{\Phi(\mathbf{Y})|\mathbf{Y} \in \mathcal{T}_j\}$.
    \STATE Compute $\pi_{\mathcal{G}}^{\mathbf{u}}(\mathbf{Y}_j)$ by projecting $\mathbf{Y}_j$ on the upper boundary of $\mathcal{G}$ (Algorithm \ref{algorithm2}).
    \IF {$\pi_{\mathcal{G}}^{\mathbf{u}}(\mathbf{Y}_j) = \mathbf{Y}_j$, i.e., $\mathbf{Y}_j \in \partial^+\mathcal{G}$}
        \STATE CBS $\mathbf{Y}^\prime = \mathbf{Y}_j$ and CBV $\Omega_j = \Phi(\mathbf{Y}_j)$.
    \ELSE
        \IF {$\Phi(\pi_{\mathcal{G}}^{\mathbf{u}}(\mathbf{Y}_j)) \geq \Omega_{j-1}$}
            \STATE $\mathbf{Y}_j^\prime = \pi_{\mathcal{G}}^{\mathbf{u}}(\mathbf{Y}_j)$ and $\Omega_j = \Phi(\pi_{\mathcal{G}}^{\mathbf{u}}(\mathbf{Y}_j))$.
        \ELSE
            \STATE $\mathbf{Y}_j^\prime = \mathbf{Y}_{j-1}^\prime$ and $\Omega_j = \Omega_{j-1}$.
        \ENDIF
        \STATE Let $\mathbf{x} = \pi_{\mathcal{G}}^{\mathbf{u}}(\mathbf{Y}_j)$ and
        $\mathcal{T}_{j+1} = (\mathcal{T}_j \setminus \mathcal{T}_*)\cup\{\mathbf{v} = \mathbf{v} + (x_t - v_t)e_t|\mathbf{v} \in \mathcal{T}_*, t\in\{1,\ldots,2K\}\}$, where $\mathcal{T}_* = \{\mathbf{v} \in \mathcal{T}_j|\mathbf{v} > \mathbf{x}\}$.
        \STATE Remove the improper vertices from $\mathcal{T}_{j+1}$.
    \ENDIF
\UNTIL {$|\Phi(\mathbf{Y}_j) - \Omega_j| \leq \epsilon$.}
\STATE $\mathbf{Y}^* = \mathbf{Y}_j^{\prime}$ is the optimal solution and corresponding $\mathbf{P}_1^*$, $\mathbf{P}_2^*$ and $\mathbf{B}^*$ is the optimal resource allocation.
\end{algorithmic}
\end{algorithm}

Via Algorithms 1--3, we determine the optimal bandwidth allocation and power allocation (BAPA) maximizing weighted sum quality of the videos of the users under total bandwidth, individual power, and individual video quality constraints (i.e., we solve the optimization problem in (\ref{Prob_Const1})).

In the numerical results presented in the next section, we demonstrate the optimal performance and identify the key tradeoffs. Additionally, we analyze the equal-bandwidth (EB) scenario in which bandwidth is equally allocated to the users, i.e., $B_k = \frac{B}{K}$, and power allocation is performed separately for each pair of full-duplex users, and provide comparisons.


\section{Numerical and Simulation Results} \label{sec:Result}
Five CIF video sequences namely \emph{Akiyo, Bus, Coastguard, Foreman} and \emph{News} are used for the simulation results \cite{Video}. Size of each frame is $352\times288$ pixels. FFMPEG is used for encoding the video sequences and GOP is set as 10. Frame rate is set as 15 frames per second. Table \ref{table_qrate} shows the parameters $a_k$ and $b_k$ that make the rate-distortion function of the five video sequences fit the quality rate model in (\ref{Qmodel1}), where the unit of $R_k$ is kbit/s. Unless mentioned explicitly, we assume that the subchannel power gain for each link is exponentially distributed with mean $Z_k = \mathbb{E}\{\gamma_k\}$. The power spectrum density of the AWGN is set to $N_0 = 10^{-6}$ W/Hz, and the channel coherence time is assumed to be 0.001 seconds. The self-interference factor at each user is set to $0.1$.

\begin{table}[h]
\begin{center}
\caption{Parameter values of the quality rate model for different video sequences} \label{table_qrate}
{
\begin{tabular}{|c|c|c|c|c|c|}
\hline
  & Akiyo & Bus & Coastguard & Foreman & News \\ \hline
$a_k$ &  5.0545 & 4.7205 & 3.5261 & 4.5006 & 5.6218  \\  \hline
$b_k$ & 17.1145 & 5.4764 & 13.8425 & 13.0780 & 10.0016 \\  \hline
\end{tabular}}
\end{center}
\end{table}

\begin{figure}
\centering
\includegraphics[width=0.4\textwidth]{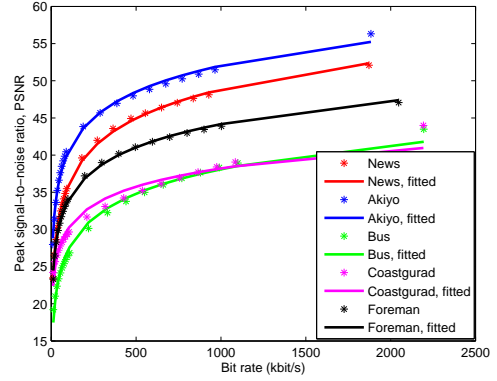}
\caption{Actual PSNR values vs. rate and fitted quality rate curves.}\label{fig:PSNR_vs Rate}
\end{figure}

Fig. \ref{fig:PSNR_vs Rate} shows the actual PSNR values as a function of the source bit rate for different video sequences, where we see that the increasing concave quality rate model fits the actual values very well. Throughout the numerical results, we assume the minimum required video quality is $Q_{i,k}^{min} = 20$dB and maximum transmission power is $P_{i,k}^{max} = 5$ for all users.

\subsection{One Pair of Full-Duplex Users}
In this section, we consider the power allocation between a single pair of full-duplex users. The bandwidth $B$ is set to $0.1$ MHz, average channel power gain is $Z_1 = 1$. $U_{1,1}$ transmits video sequence  \emph{Bus} to $U_{2,1}$, while $U_{2,1}$ transmits video sequence \emph{Coastguard} to $U_{1,1}$, with the corresponding parameters $(a_{1,1} = 4.7205$, $b_{1, 1} = 5.4764)$, and $(a_{2, 1} = 3.5261$, $b_{2, 1} = 13.8425)$ from Table I.

\subsubsection{The Impact of the QoS Exponent on Multimedia Quality}

\begin{figure}
\centering
\begin{subfigure}[b]{0.35\textwidth}
\includegraphics[width=\textwidth]{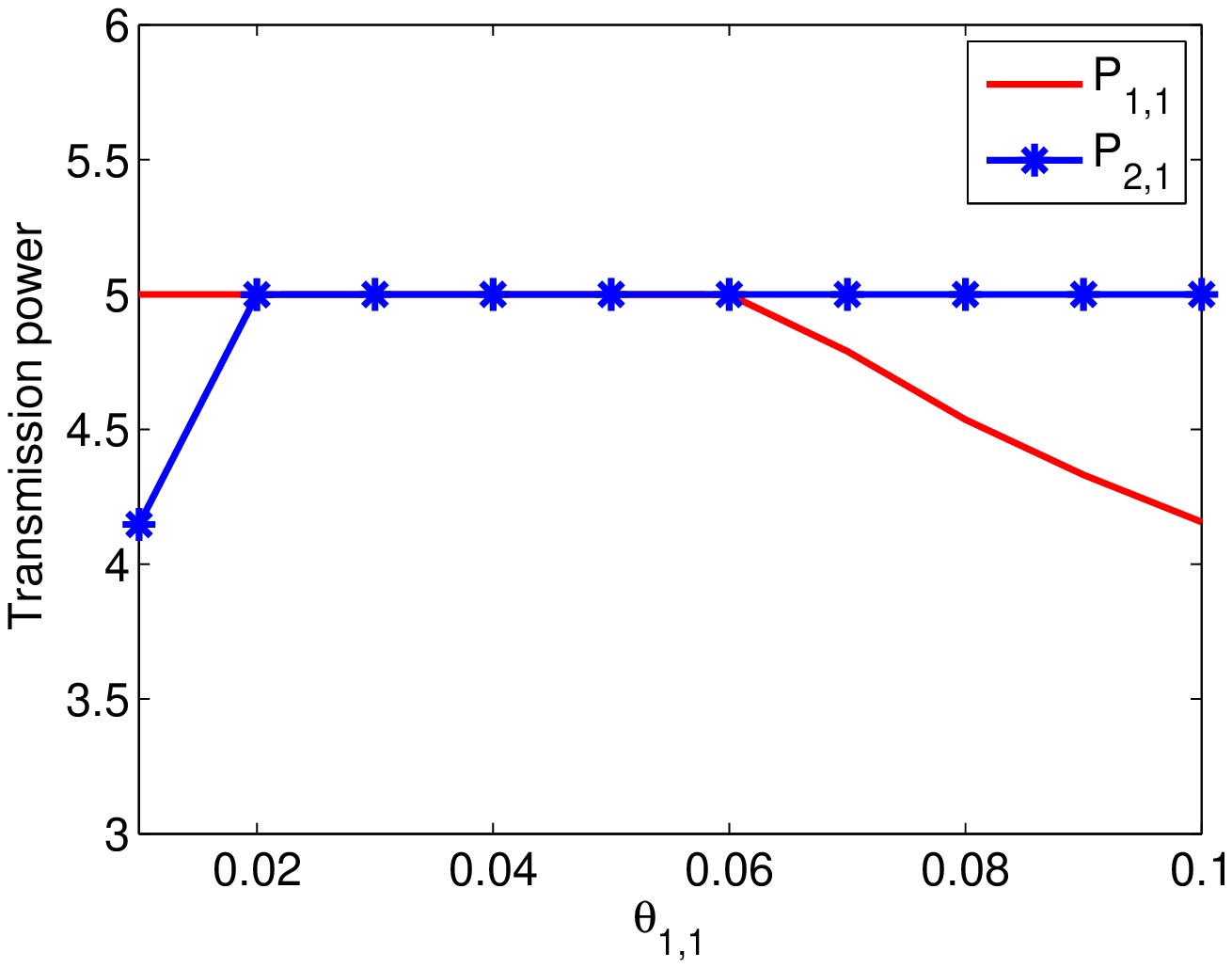}
\caption{\small{}}\label{fig:PowersvsTheta1}
\end{subfigure}
\begin{subfigure}[b]{0.35\textwidth}
\includegraphics[width=\textwidth]{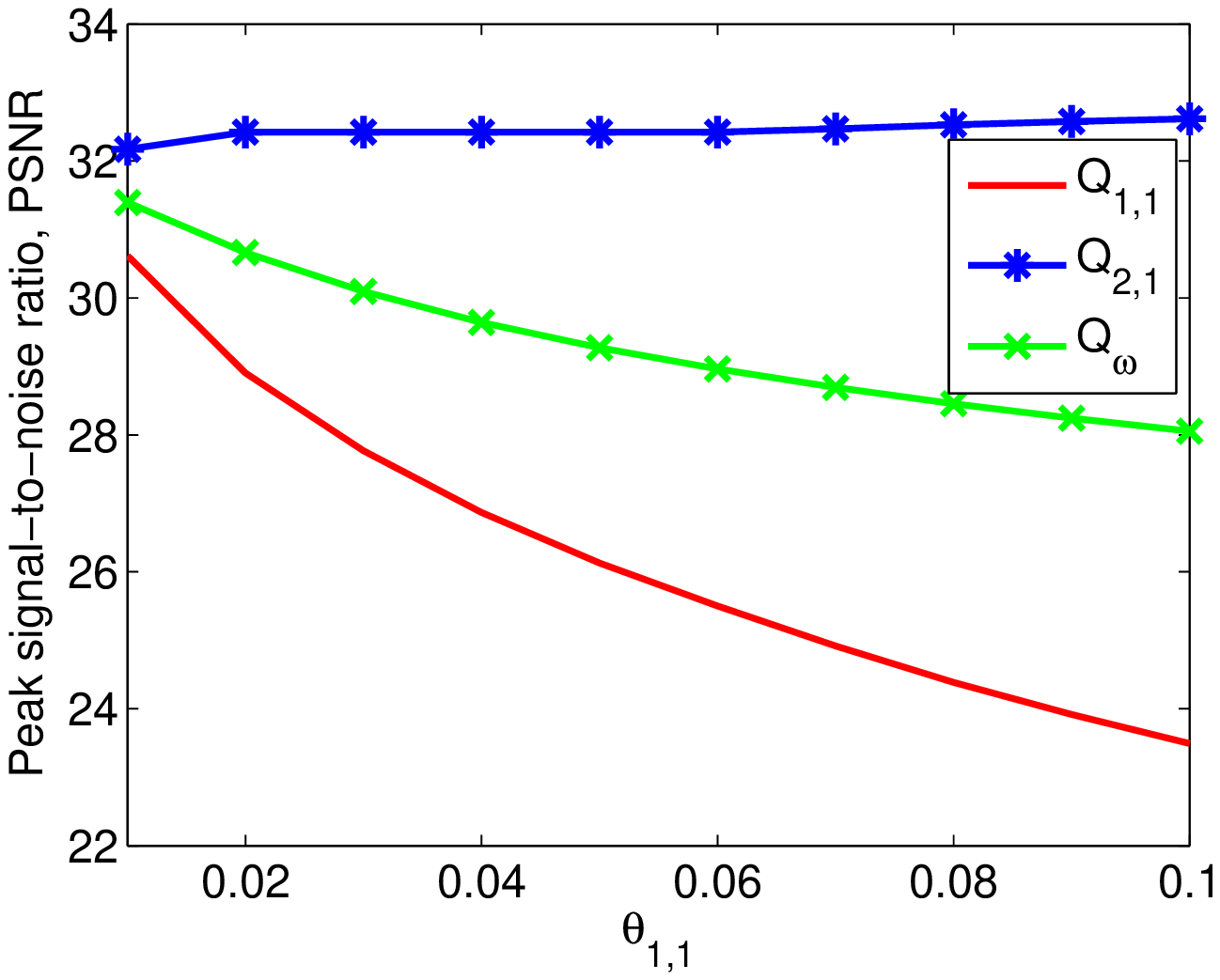}
\caption{\small{}}\label{fig:PSNRvsTheta1}
\end{subfigure}
\caption{(a) Optimal power allocation and (b) the corresponding quality $Q$ (or equivalently PSNR) of video sequences as a function of $\theta_{1,1}$.}\label{fig:PSNRPowervsTheta1}
\end{figure}

In Fig. \ref{fig:PSNRPowervsTheta1}, we set $\omega_{1,1} = \omega_{2,1} = 0.5$ (meaning that two videos are equally weighted), and increase the value of $\theta_{1, 1}$ (the QoS exponent of user $U_{1,1}$) from $0.01$ to $0.1$ while keeping $\theta_{2, 1} = 0.01$. Note that increased $\theta_{1,1}$ implies that more stringent delay constraints are imposed on the video transmission of $U_{1,1}$. Fig. \ref{fig:PowersvsTheta1} plots the power allocated to the users as $\theta_{1, 1}$ increases. Since quality parameter $a_{1,1}$ of \emph{Bus} video is greater than $a_{2,1}$ of \emph{Coastguard} video, quality $Q_{1,1}$ of the \emph{Bus} video increases faster than $Q_{2,1}$ of the \emph{Coastguard} video as the transmission power and correspondingly the arrival rate $R$ grow, according to the logarithmic model in (\ref{Qmodel1}). Therefore, initially when $\theta_{1, 1} = \theta_{2, 1} = 0.01$ and $U_{1,1}$ and $U_{2,1}$ are subject to the same delay constraint, $U_{1,1}$ transmits at the peak power level in a greedy fashion to maximize the sum video quality, while $U_{2,1}$ uses less power.

As $\theta_{1, 1}$ increases, more stringent delay constraints are imposed on user $U_{1,1}$ and the arrival rate $R_{1, 1}$ of the \emph{Bus} video is reduced to avoid delay violations. Consequently, the video quality $Q_{1,1}$ (or equivalently the PSNR of the video) starts diminishing as seen in Fig. \ref{fig:PSNRvsTheta1}. Eventually, when $\theta_{1,1}$ exceeds 0.06, the lower arrival rates can be supported by smaller transmission power and $P_{1,1}$ is reduced as observed in Fig. \ref{fig:PowersvsTheta1}. In the meantime, we notice that quality $Q_{2,1}$ of \emph{Costguard} video slightly increases due to increased transmission power $P_{2,1}$ at $U_{2,1}$ and smaller self-interference at $U_{1,1}$ (because of smaller transmission power $P_{1,1}$). However, since the drop in $Q_{1,1}$ is more significant, the weighted sum quality $Q_{\omega}$ is seen to decrease in Fig. \ref{fig:PSNRvsTheta1}. Finally, it is interesting to note that, as predicted by Theorem \ref{theo:min_bandw} and discussed subsequently, at least power value is at the maximum level of 5, i.e., $P_{1,1} = 5$ or $P_{2,1} = 5$, for any given value of $\theta_{1,1}$ in Fig. \ref{fig:PowersvsTheta1}.


\begin{figure}
\centering
\begin{subfigure}[b]{0.35\textwidth}
\includegraphics[width=\textwidth]{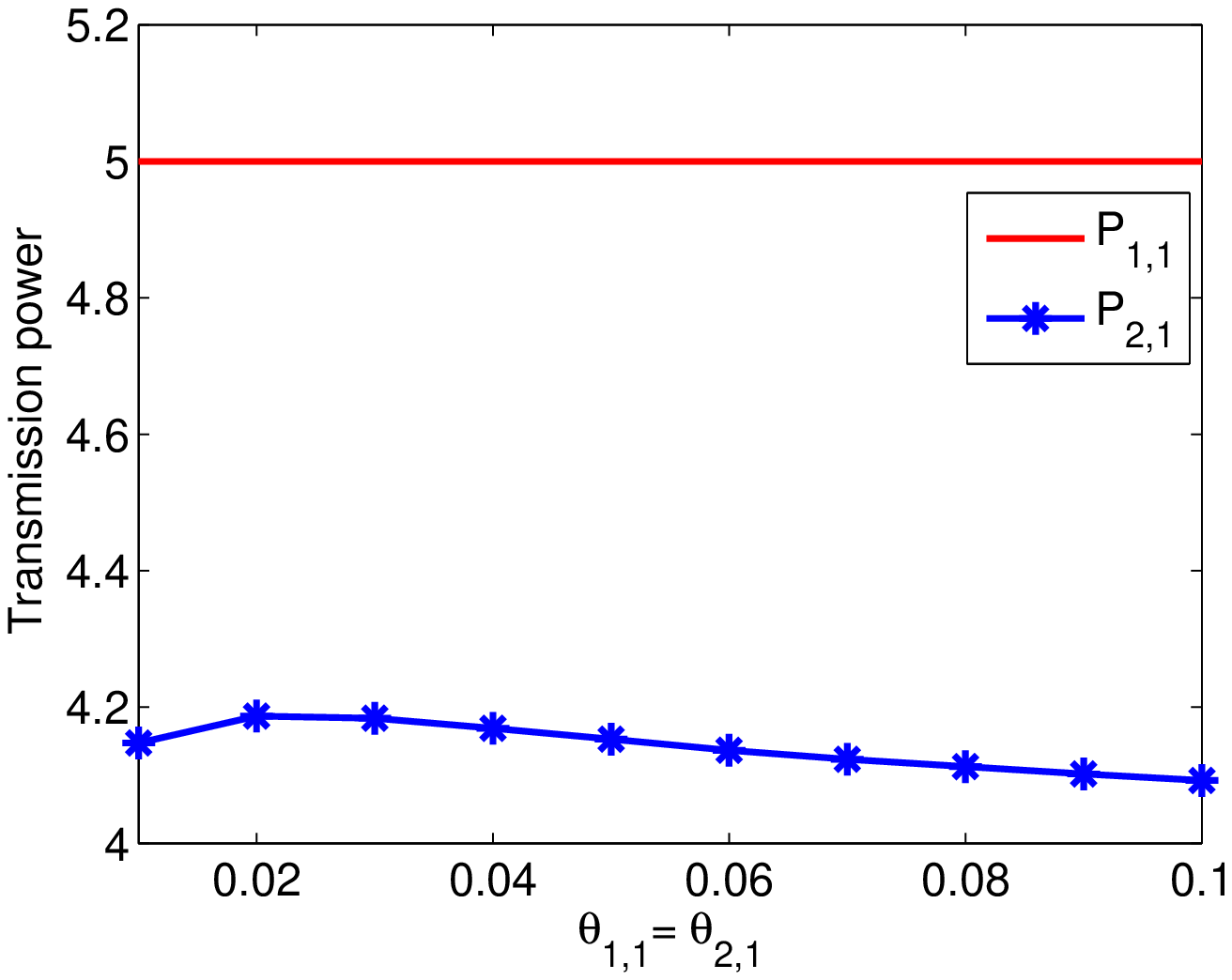}
\caption{\small{}}\label{fig:PowervsTheta}
\end{subfigure}
\begin{subfigure}[b]{0.35\textwidth}
\includegraphics[width=\textwidth]{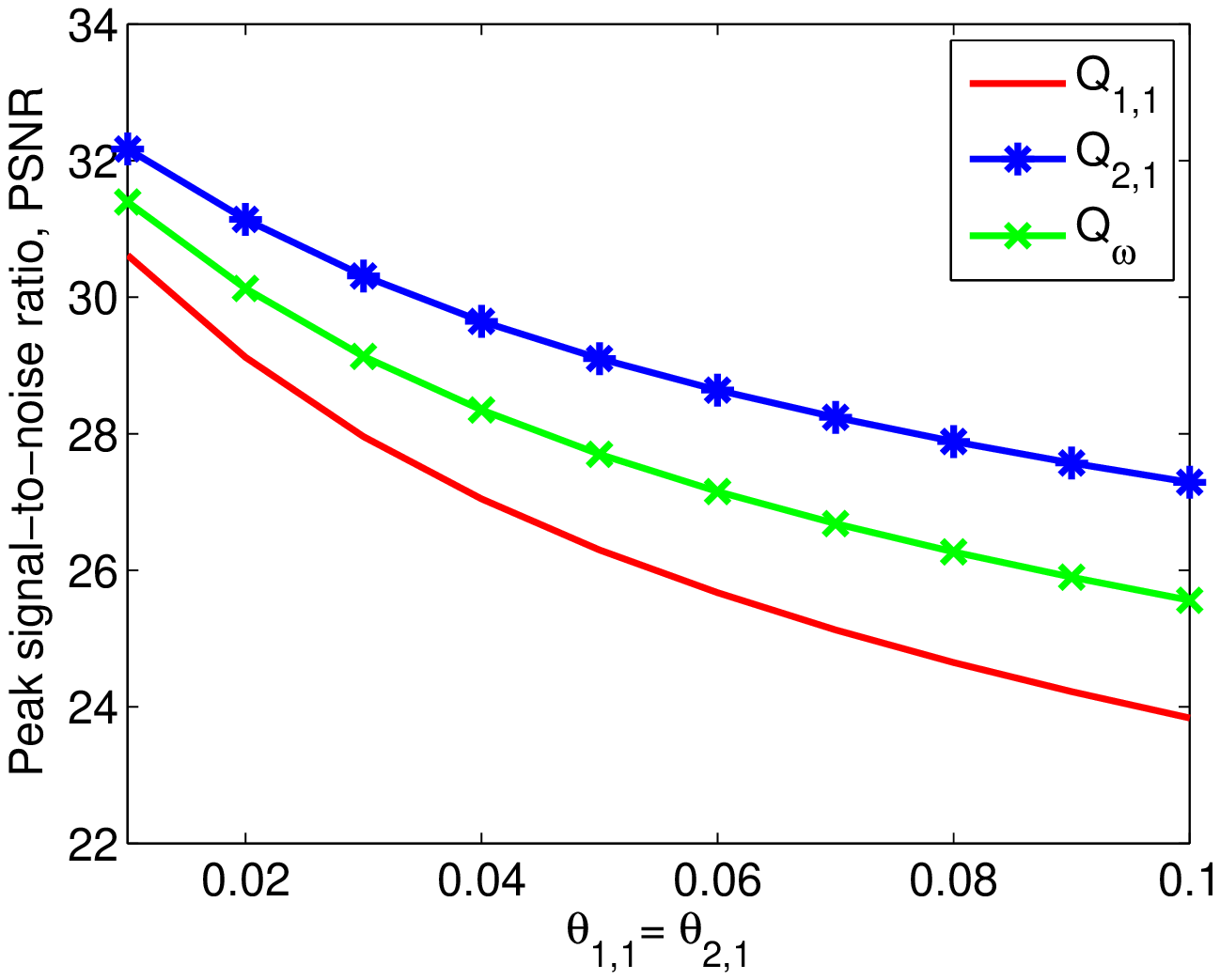}
\caption{\small{}}\label{fig:PSNRvsTheta}
\end{subfigure}
\caption{(a) Optimal power allocation and (b) the corresponding quality $Q$ (or equivalently PSNR) of video sequences as a function of $\theta_{1,1} = \theta_{2,1}$.}\label{fig:PSNRPowervsTheta}
\end{figure}

In Fig. \ref{fig:PSNRPowervsTheta}, both $\theta_{1, 1}$ and $\theta_{2, 1}$ increase from $0.01$ to $0.1$ together. Since $U_{1,1}$ and $U_{2,1}$ now all the time operate under the same QoS constraints while transmitting different video sequences, Fig. \ref{fig:PowervsTheta} demonstrates that $P_{1,1}$ is always greater than $P_{2,1}$ due to, as discussed above, the impact of video quality parameters, or more specifically due to having $a_{1,1} > a_{2,1}$. Fig. \ref{fig:PSNRvsTheta} shows that both $Q_{1,1}$ and $Q_{2,1}$ decrease as both $\theta_{1, 1}$ and $\theta_{2, 1}$ increase. That is because larger $\theta_{1, 1}$ and $\theta_{2, 1}$ lead to smaller source rates $R_{1,1}$ and $R_{2,1}$, which in turn reduce the video quality.

\subsubsection{The Impact of Weights on Multimedia Quality}
\begin{figure}
\centering
\begin{subfigure}[b]{0.35\textwidth}
\includegraphics[width=\textwidth]{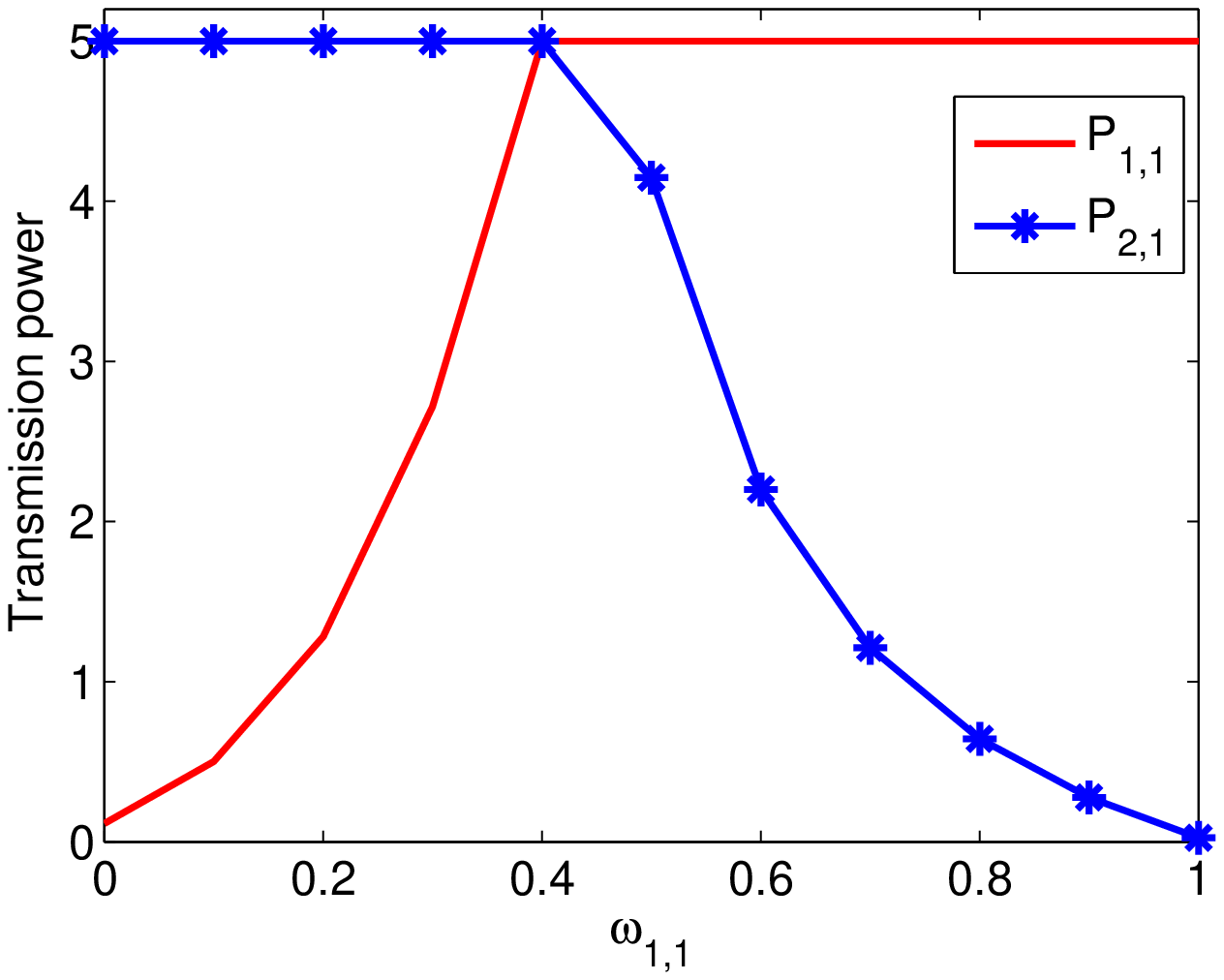}
\caption{\small{}}\label{fig:Powersvsomega1}
\end{subfigure}
\begin{subfigure}[b]{0.35\textwidth}
\includegraphics[width=\textwidth]{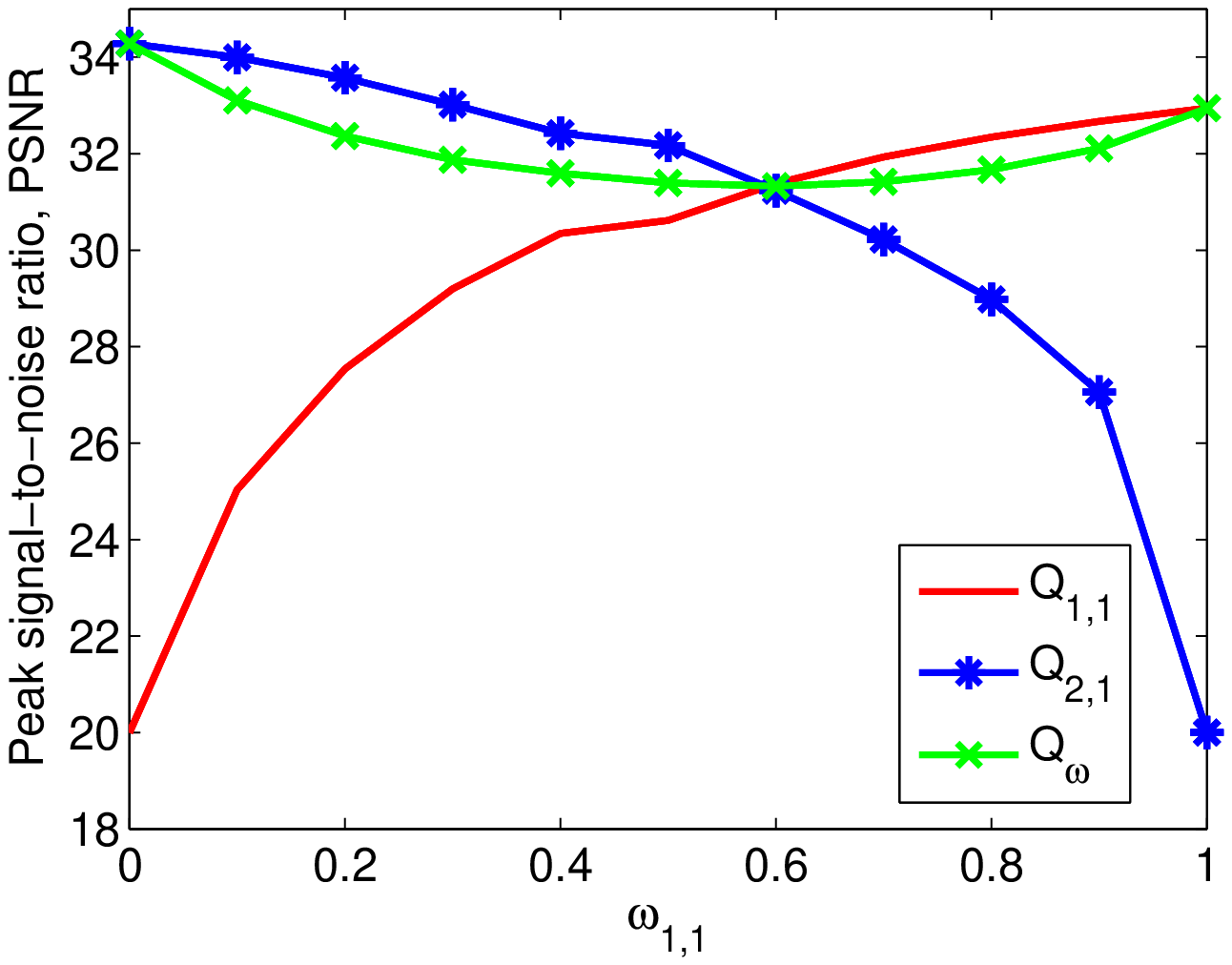}
\caption{\small{}}\label{fig:PSNRvsomega1}
\end{subfigure}
\caption{(a) Optimal power allocation and (b) the corresponding quality $Q$ (or equivalently PSNR) of video sequences as a function of $\omega_{1,1}$.}\label{fig:PSNRPowervsomega1}
\end{figure}

Now, we set $\theta_{1,1} = \theta_{2,1} = 0.01$, and increase the weight $\omega_{1,1}$ from $0$ to $1$ while keeping $\omega_{1,1} + \omega_{2,1} = 1$.  Hence, the weight of user $U_{1,1}$ gradually increases in the weighted sum quality maximization in (\ref{Prob_Const1}). Fig. \ref{fig:Powersvsomega1} shows that, as expected, $P_{1,1}$ grows and reaches the peak value as $\omega_{1,1}$ increases due to higher emphasis on the quality $Q_{1,1}$. At the same time, $P_{2,1}$ starts diminishing when $\omega_{1,1}$ increases beyond 0.4 and hence $\omega_{2,1}$ drops below 0.6. Fig. \ref{fig:PSNRvsomega1} plots the corresponding qualities of the video sequences. Following similar trends as in the power curves, $Q_{1,1}$ improves whereas $Q_{2,1}$ is reduced. Finally, we note that we have $Q_{1,1} = 20$dB when $\omega_{1,1} = 0$, and $Q_{2,1} = 20$dB  when $\omega_{1,1} = 1$ due to the fact that a minimum quality of $20$dB is imposed on both video transmissions.

\subsection{Two Pairs of Full-Duplex Users}
In this section, we consider bandwidth and power allocation for two pairs of full-duplex users. The total bandwidth $B$ is set to $0.2$ MHz, and the average channel power gains are $Z_1 = 1$ between first pair of users and $Z_2 = 3$  between the second pair of users. $U_{1,1}$ and $U_{1,2}$ transmit the same video sequence \emph{Bus} to $U_{2,1}$ and $U_{2,2}$, respectively. And video sequence \emph{Coastguard} is transmitted to $U_{1,1}$ and $U_{1,2}$ by $U_{2,1}$ and $U_{2,2}$ respectively. For these video sequences, we have $a_{1,1} = a_{1,2} = 4.7205$ and $b_{1, 1} = b_{1,2} = 5.4764$, $a_{2, 1} = a_{2,2} = 3.5261$ and $b_{2, 1} = b_{2,2} = 13.8425$.

\subsubsection{The Impact of the QoS Exponent on Multimedia Quality}
\begin{figure*}
\centering
\begin{subfigure}[b]{0.3\textwidth}
\includegraphics[width=\textwidth]{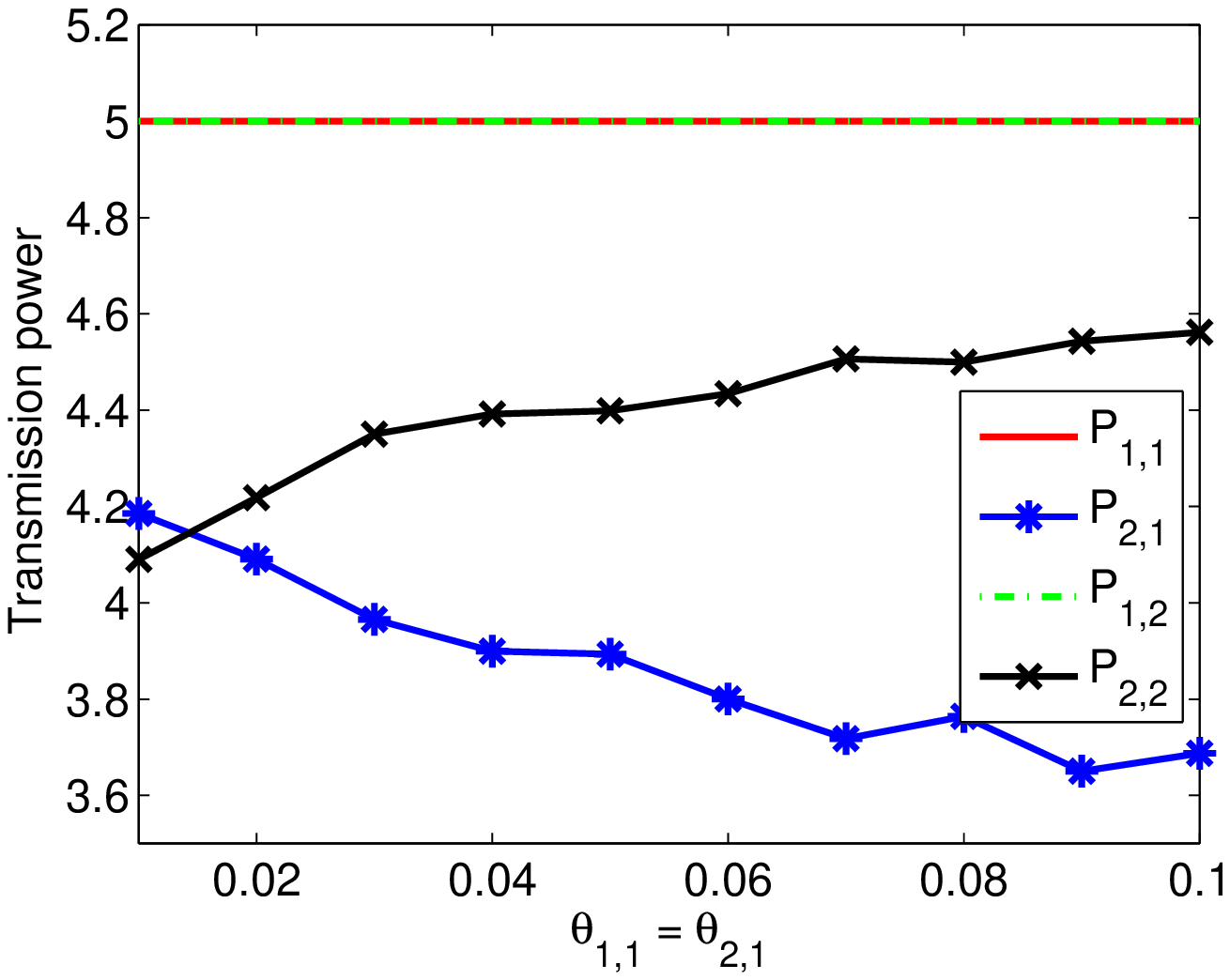}
\caption{\small{}}\label{fig:Powersvstheta_2p}
\end{subfigure}
\begin{subfigure}[b]{0.3\textwidth}
\includegraphics[width=\textwidth]{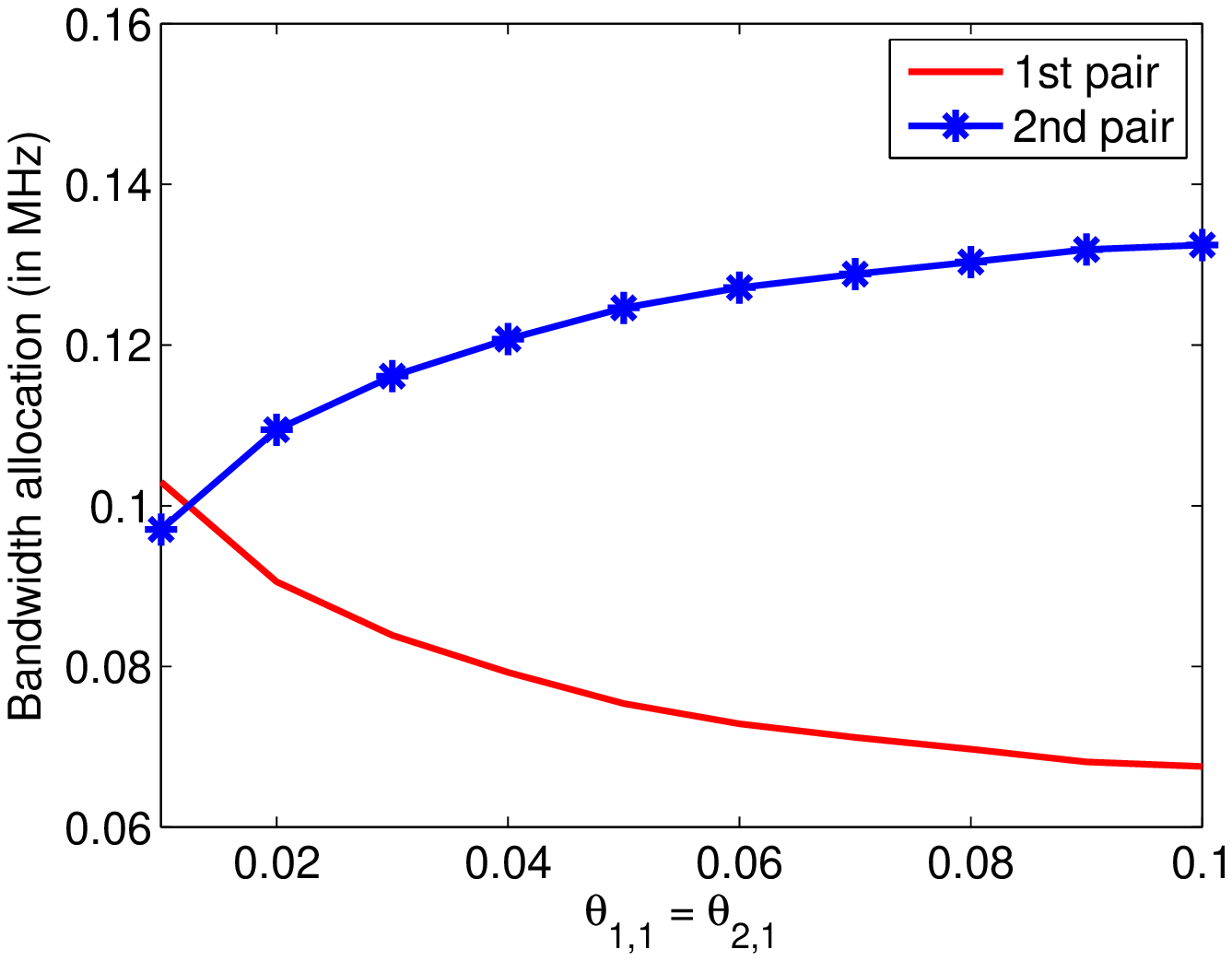}
\caption{\small{}}\label{fig:Bandvstheta_2p}
\end{subfigure}
\begin{subfigure}[b]{0.3\textwidth}
\includegraphics[width=\textwidth]{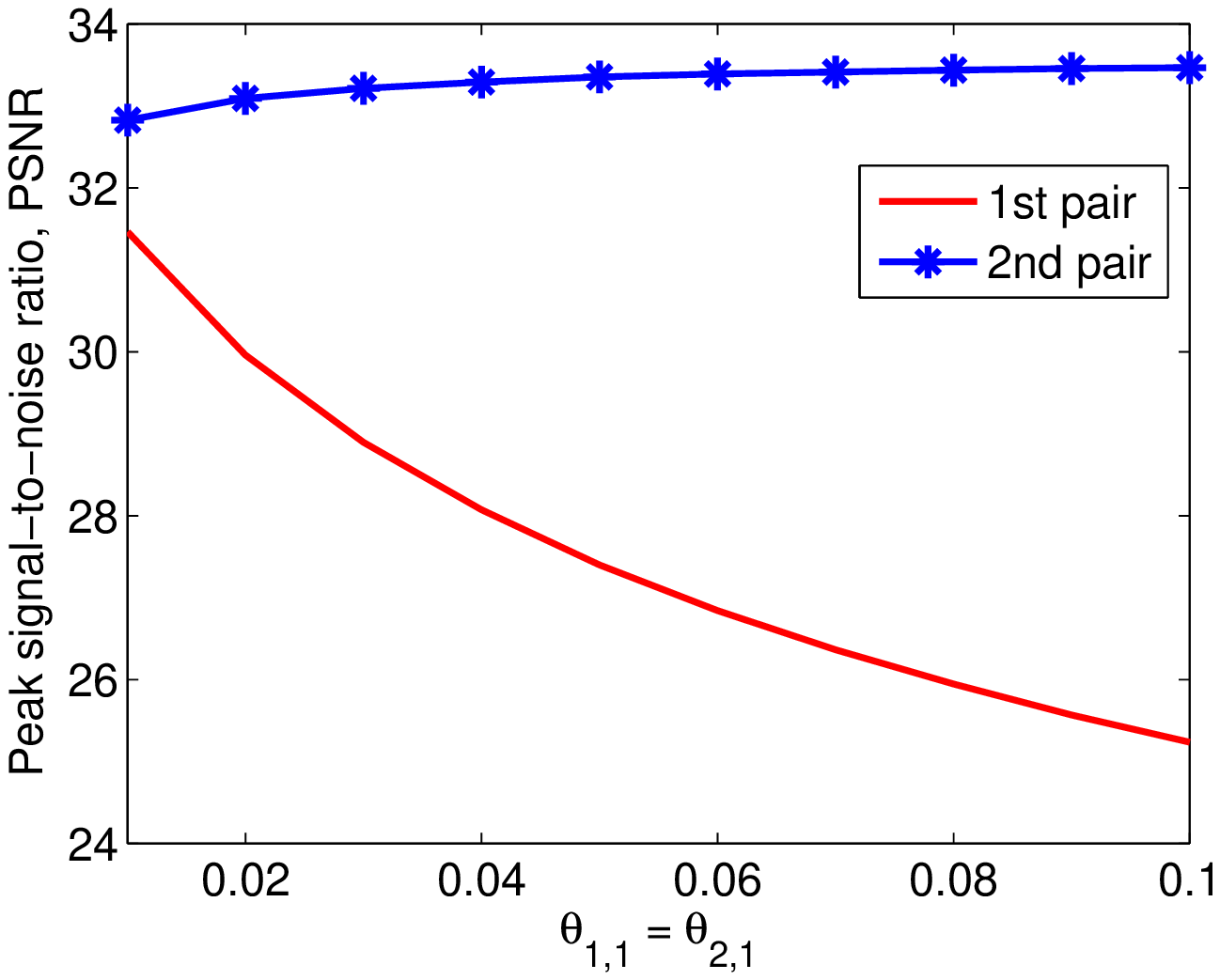}
\caption{\small{}}\label{fig:PSNRvstheta_2p}
\end{subfigure}
\caption{(a) Optimal power allocation, (b) optimal bandwidth allocation, and  (c) the corresponding quality of video sequences as $\theta_{1,1} = \theta_{2,1}$ increase.}\label{fig:PSNRPowervstheta_2p}
\end{figure*}

In this subsection, we initially set $\omega_{1,1} = \omega_{2,1} = \omega_{1,2} = \omega_{2,2} = 0.25$, and increase the values of the QoS exponents of the first pair of users $\theta_{1, 1}$ and  $\theta_{2, 1}$ from $0.01$ to $0.1$ together (i.e., $\theta_{1, 1} = \theta_{2, 1}$) while keeping the QoS exponents of the second pair of users at $\theta_{1, 2} = \theta_{2, 2} = 0.01$. Fig. \ref{fig:Powersvstheta_2p} and Fig. \ref{fig:Bandvstheta_2p} show the results of the optimal power and bandwidth allocation as a function of $\theta_{1,1}=\theta_{2,1}$. Note that as QoS exponents $\theta_{1,1} = \theta_{2,1}$ increase (hence more stringent QoS constraints are imposed), lower arrival rates are supported and the quality of the video sequences of the first pair of users degrades. With this, bandwidth allocated to the first pair of users is reduced as noticed in Fig. \ref{fig:Bandvstheta_2p}. Due to similar reasons (regarding the video quality parameters) as discussed in the case of one pair of full-duplex users (i.e., $a_{1,k} > a_{2,k}$ for $k = 1,2$), $P_{1,1}$ and $P_{1,2}$ are always at their maximum levels. We also observe that as $\theta_{1,1}=\theta_{2,1}$ increase, $P_{2,1}$ diminishes whereas $P_{2,2}$ grows. These are due to the facts that the bandwidth allocated to the link between $U_{1,1}$ and $U_{2,1}$ decreases while the bandwidth allocated to the link between $U_{1,2}$ and $U_{2,2}$ increases. Hence, an opportunistic strategy is employed and more power is allocated to the link with more bandwidth.
Fig. \ref{fig:PSNRvstheta_2p} demonstrates that the average PSNR value of first pair of video sequences degrades due to increasing QoS exponents and smaller bandwidth.

\begin{figure}
\centering
\includegraphics[width=0.35\textwidth]{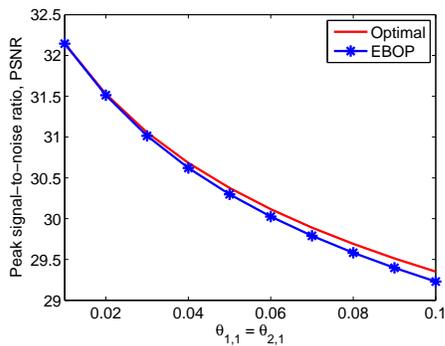}
\caption{Quality of video sequences as a function of $\theta_{1,1} = \theta_{2,1}$. Both optimal and equal bandwidth allocation are considered.}\label{fig:Scheme_2p}
\end{figure}

Fig. \ref{fig:Scheme_2p} plots the weighted sum quality of video sequences assuming optimal and also equal bandwidth allocation. In both cases, power is optimally allocated. We note that the equal bandwidth optimal power (EBOP) allocation scheme provides a performance close to that of the optimal bandwidth and power allocation scheme, but the gap widens as $\theta_{1,1} = \theta_{2,1}$ increase. Below, we will demonstrate that performance gap expands further when we have unequal weights.

\subsubsection{The Impact of Weights on Multimedia Quality}
\begin{figure*}
\centering
\begin{subfigure}[b]{0.3\textwidth}
\includegraphics[width=\textwidth]{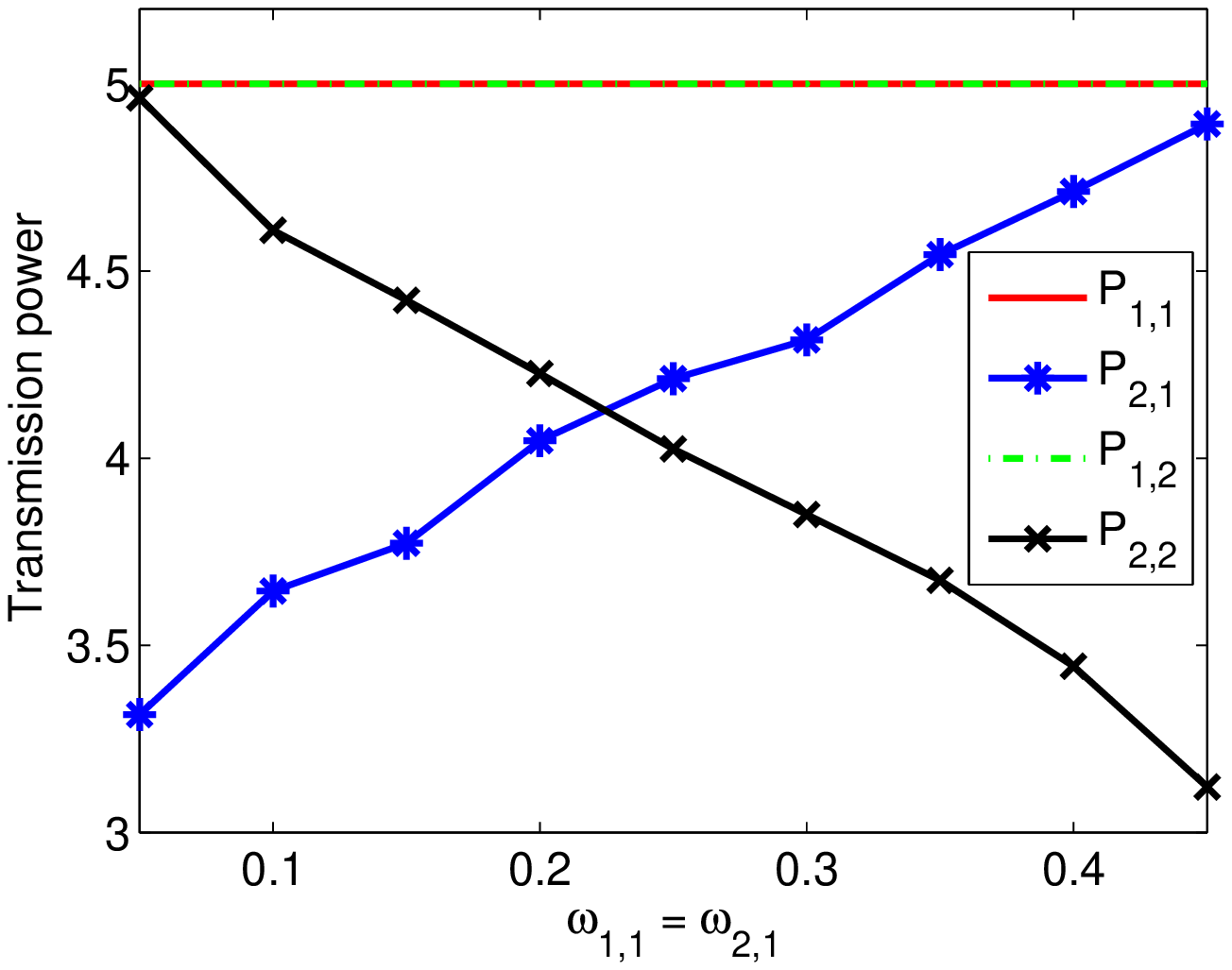}
\caption{\small{}}\label{fig:Powers2pvsomega}
\end{subfigure}
\begin{subfigure}[b]{0.3\textwidth}
\includegraphics[width=\textwidth]{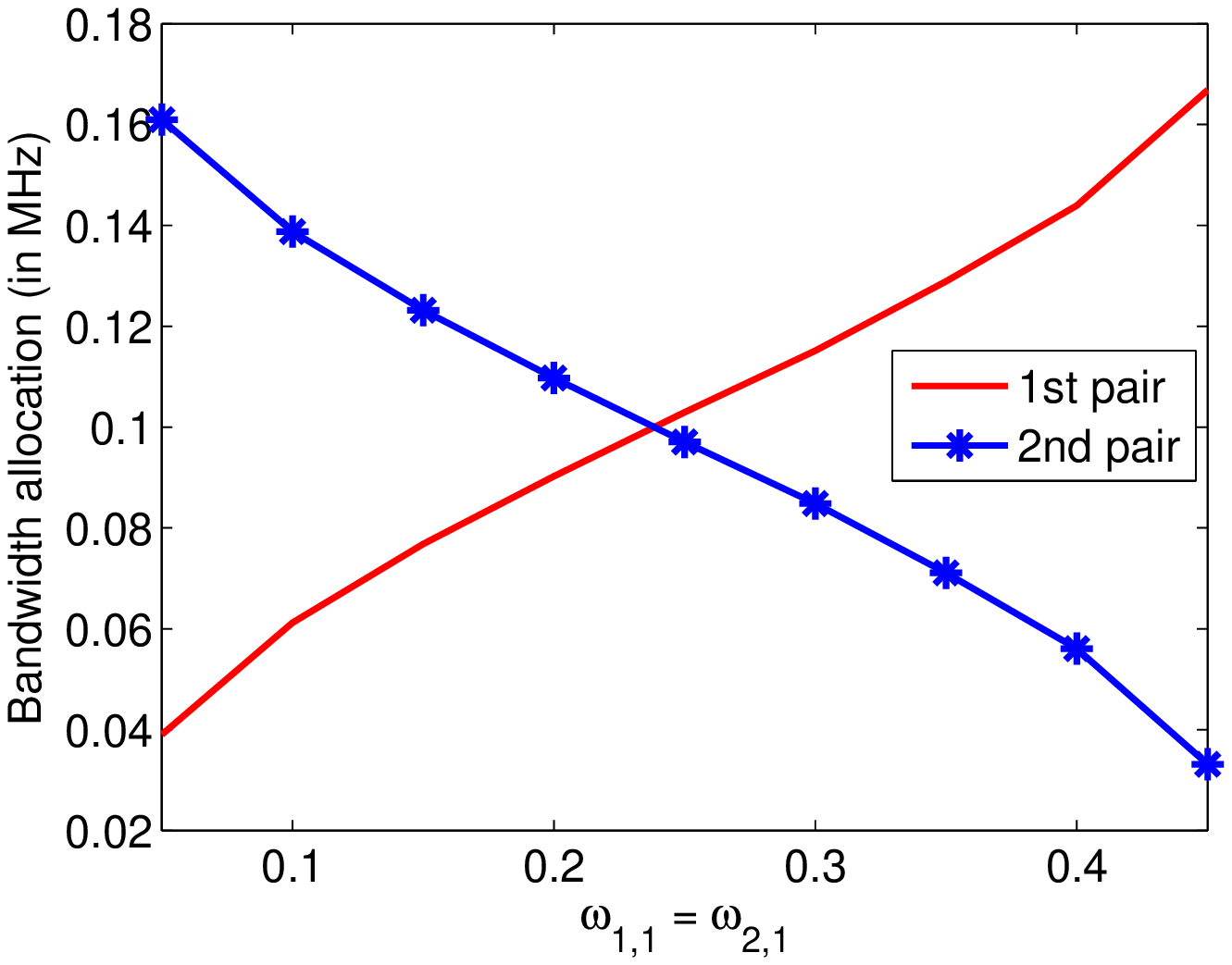}
\caption{\small{}}\label{fig:Band2pvsomega}
\end{subfigure}
\begin{subfigure}[b]{0.3\textwidth}
\includegraphics[width=\textwidth]{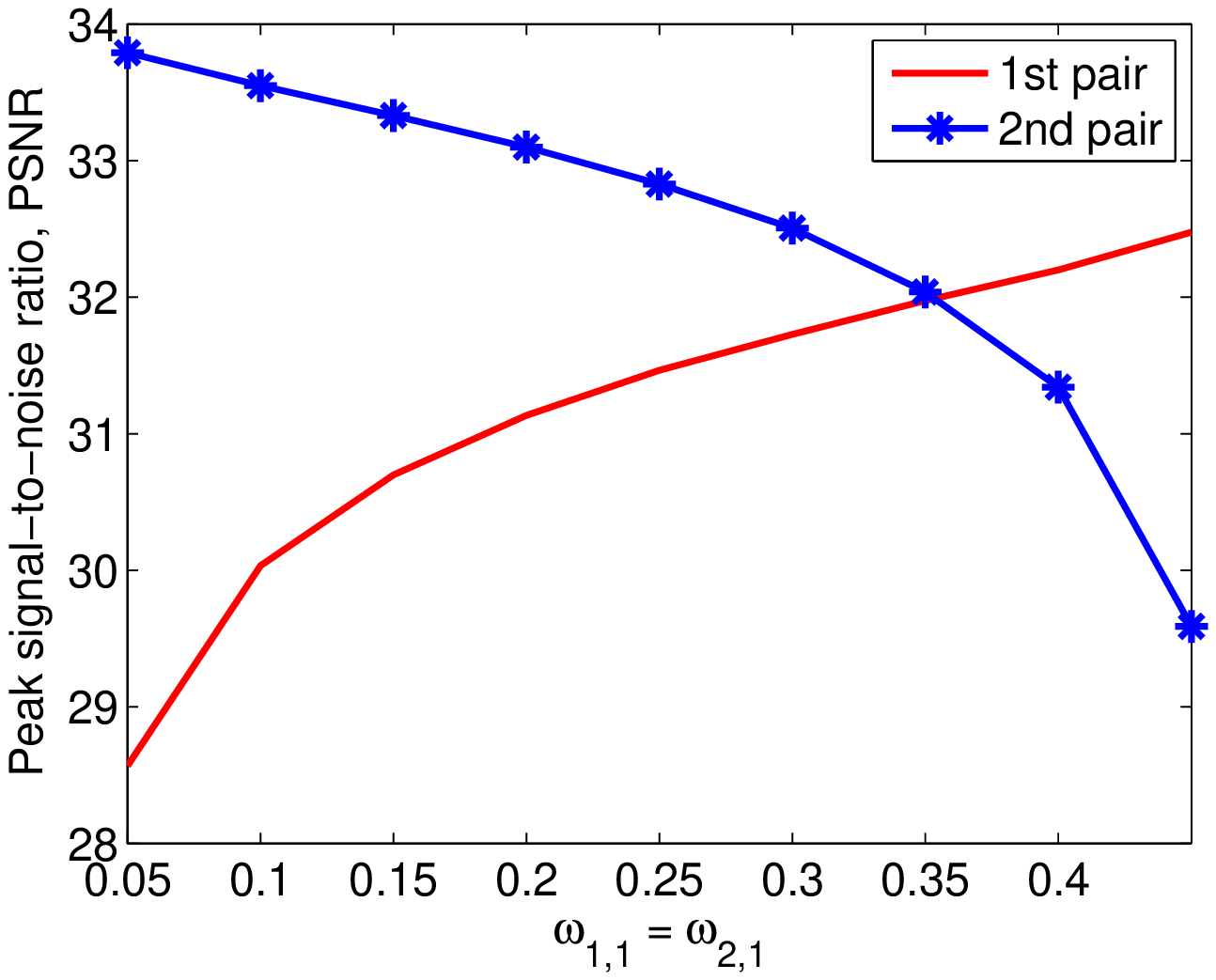}
\caption{\small{}}\label{fig:PSNR2pairvsomega}
\end{subfigure}
\caption{(a) Optimal power allocation, (b) Optimal bandwidth allocation, and (c) the corresponding quality of video sequences as a function of $\omega_{1,1} = \omega_{2,1}$.}\label{fig:PSNRPowervsomega_2p}
\end{figure*}

Fig. \ref{fig:PSNRPowervsomega_2p} shows the optimal bandwidth and power allocation and the corresponding quality of video sequences as the weights $\omega_{1,1} = \omega_{2,1}$ vary from $0.05$ to $0.45$. We also assume that $\omega_{1, 2} = \omega_{2, 2}$ while keeping the sum of all weights equal to 1. Fig. \ref{fig:Band2pvsomega} indicates that bandwidth $B_1$ allocated to the first pair of users increases with increasing $\omega_{1, 1} = \omega_{2,1}$ since growing emphasis is given to the quality of the video sequences transmitted between first pair of users. Consequently, the bandwidth allocated to the link between second pair of users $U_{1,2}$ and $U_{2,2}$ decreases. Since $\omega_{1, 1} = \omega_{2,1}$ and $a_{1,1}>a_{2,1}$, $U_{1,1}$ always transmits the video sequence at the maximum transmission power level. Due to the same reason, $P_{1, 2}$ always attains the maximum level. Again, due to the optimality of the opportunistic approach, $P_{2,1}$ increases as $B_1$ gets larger, whereas $P_{2,2}$ diminishes as $B_2$ becomes smaller. Correspondingly, Fig. \ref{fig:PSNR2pairvsomega} demonstrates that the average PSNR values $Q_{1,1}$ and $Q_{2,1}$ improve as higher weights  $\omega_{1, 1} = \omega_{2,1}$ are given to the video communication between the first pair of users, while the average PSNR values $Q_{1,2}$ and $Q_{2,2}$ are lowered.

\begin{figure}
\centering
\includegraphics[width=0.35\textwidth]{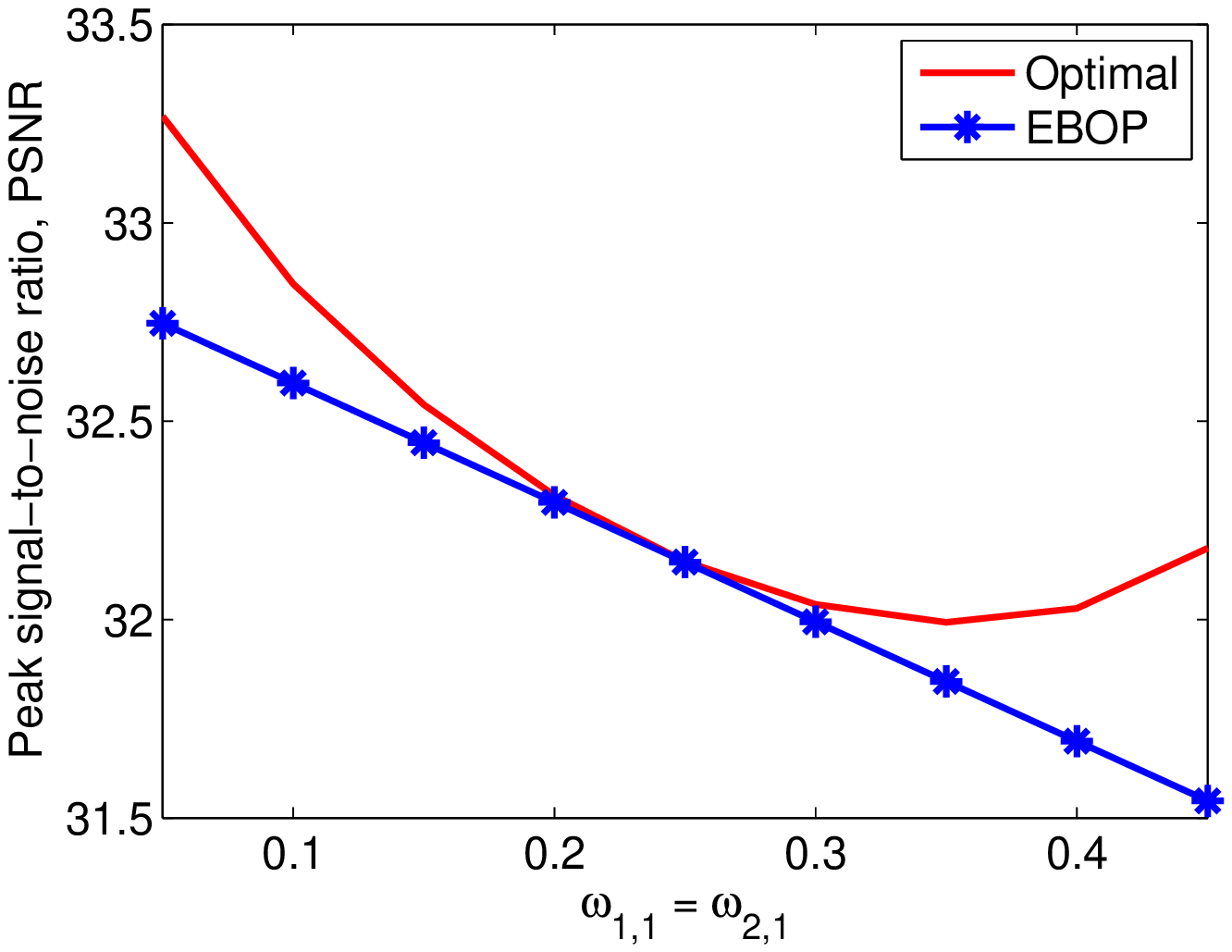}
\caption{Quality of video sequences as a function of $\omega_{1,1} = \omega_{2,1}$. Both optimal and equal bandwidth allocation are considered.}\label{fig:Scheme_2p_omega}
\end{figure}

Fig. \ref{fig:Scheme_2p_omega} shows the weighted sum quality of video sequences again considering optimal and equal bandwidth allocation schemes. As expected, the optimal bandwidth and power allocation scheme outperforms the case in which bandwidth is equally allocated among the pairs of users and power is allocated optimally (i.e., EBOP scheme). The performance gap is smallest when the weights are all equal (i.e.,  $\omega_{1,1} = \omega_{2,1} = \omega_{1,2} = \omega_{2,2} = 0.25$), and the gap grows as the difference in the weights increases.

\subsection{More than Two Pairs of Full-Duplex Users}
In this subsection, we apply our optimal resource allocation  algorithms to cases in which there are more than two pairs of full-duplex users   Table \ref{table_3pairs} provides results on the optimal bandwidth and power allocation and the resulting video qualities when there 3 pairs of users.
In these results, it is assumed that $\omega_{1,1} = \omega_{2, 1} = 0.05$, $\omega_{1,2} = \omega_{2, 2} = 0.3$ and $\omega_{1,3} = \omega_{2, 3} = 0.15$. Moreover, we set $\theta_{1,1} = \theta_{2, 1} = 0.1$, $\theta_{1,2} = \theta_{2, 2} = 0.07$ and $\theta_{1,3} = \theta_{2, 3} = 0.04$.
Overall, optimal bandwidth and power allocation leads to a weighted sum quality of $33.9269$dB. We notice that since the weights $\omega_{1,2}$ and $\omega_{2,2}$ of the second pair of users are the largest, most bandwidth (out of a total bandwidth of $B = 0.3$MHz $= 300$kHz) is allocated to these users. Also, it is interesting to note that due to the need to control the self-interference, several power levels are less than the maximum allowed peak power level of 5 (while at least one power value is at the peak level), as also noted in the previous cases.


\begin{table}[h]
\begin{center}
\caption{Performance with 3 pairs of full-duplex users} \label{table_3pairs}
{
\begin{tabular}{|c|c|c|c|c|c|}
\hline
$k$ & $P_{1,k}$ & $P_{2,k}$ & $B_k$ & $Q_{1,k}$ & $Q_{2,k}$ \\ \hline
1 & 5 & 3.8971 & 51.626 & 23.2390 & 26.7099 \\ \cline{1-6}
2 & 5 & 4.0473 & 150.691& 34.5854 & 38.8709 \\ \cline{1-6} 3 & 5 & 4.3400 & 97.683 & 28.1572 & 34.4601 \\ \hline
\end{tabular}}
\end{center}
\end{table}

Table \ref{table_4pairs} shows the performances of video transmissions between 4 pairs of full-duplex users again considering optimal bandwidth and power allocation with $\omega_{1,1} = \omega_{2, 1} = 0.05$, $\omega_{1,2} = \omega_{2, 2} = 0.2$, $\omega_{1,3} = \omega_{2, 3} = 0.05$ and $\omega_{1,4} = \omega_{2, 4} = 0.2$. The total bandwidth is $B = 0.4$MHz $= 400$kHz.
It is further assumed that $\theta_{1,1} = \theta_{2, 1} = 0.1$, $\theta_{1,2} = \theta_{2, 2} = 0.07$, $\theta_{1,3} = \theta_{2, 3} = 0.04$ and $\theta_{1,4} = \theta_{2, 4} = 0.01$. The weighted sum quality of video sequences achieved with optimal allocations is $36.8243$dB.

\begin{table}[h]
\begin{center}
\caption{Performance with 4 pairs of full-duplex users} \label{table_4pairs}
{
\begin{tabular}{|c|c|c|c|c|c|}
\hline
$k$ & $P_{1,k}$ & $P_{2,k}$ & $B_k$ & $Q_{1,k}$ & $Q_{2,k}$\\ \hline
1 & 5 & 3.7464 & 26.720 & 22.4085 & 25.9285\\ \cline{1-6}
2 & 5 & 4.9929 & 146.759& 34.4014 & 39.0498  \\ \cline{1-6}
3 & 5 & 4.3079 & 36.258 & 26.7723 & 33.0397  \\ \cline{1-6}
4 & 5 & 4.9990 & 190.263 & 43.6321 & 40.0009 \\ \hline
\end{tabular}}
\end{center}
\end{table}

\section{Conclusion} \label{sec:Conclusion}
In this paper, we have addressed the maximization of the weighted sum quality of received video sequences under total bandwidth, minimum video quality, maximum transmission power, and delay QoS constraints in a full-duplex wireless model. LTPRS model is employed as the full-duplex model and the self-interference is measured by multiplying a self-interference factor with the transmission power. We have reformulated the original nonconvex optimization problem as a monotonic optimization problem, and developed algorithms to determine the optimal bandwidth and power allocation levels in an efficient manner using this framework.

We have gleaned several practical insights from our analysis. We have shown that larger values of the QoS exponent $\theta$ lead to lower PSNR levels since more stringent delay constraints result in smaller video rates, lowering the quality. We have also demonstrated that the user with a larger $\theta$ is allocated smaller transmission power and bandwidth. We have seen that video quality parameters have influence on optimal resource allocation policies, e.g., if the video quality increases faster with increased source rate (i.e., $a_{i,k}$ is larger for a video sequence), transmission power is higher. Furthermore, we have noted that weights $\omega$ in quality maximization play an important role and users whose video qualities are assigned a large weight are allocated more resources and attain higher PSNR values for the reconstructed video. We have also shown that optimal bandwidth and power allocation has better performance than the equal bandwidth and optimal power (EBOP) allocation scheme, and the performance gap widens as the weight differences among the transmitted videos grow.

\appendix
\subsection{Proof of Theorem \ref{theo:min_bandw}} \label{Proof:1}
Let us assumer $P_1 \leq P^{\max}$ and $P_2 \leq P^{\max}$, and consider the function
\begin{align} \label{eq:V1function}
V_1(P_1, P_2, B) = \left(\mathbb{E}_{\gamma}\left\{ e^{-\theta B T_c \log\left(1+\frac{P_1 \gamma}{N_0 B + \mu P_2}\right)}\right\}\right)^{-1}.
\end{align}
We first show that $V_1$ is maximized if $P_1 = P^{\max}$ or $P_2 = P^{\max}$. Hence, at least one power value should be at the maximum level. Consider two power values strictly less than the maximum level, i.e., $P_1 < P^{\max}$ and $P_2 < P^{\max}$. Then, there exists some $\tau > 1$ such that $\tau P_1 \leq P^{\max}$ and $\tau P_2 \leq P^{\max}$. Then, considering the fraction in the exponent in (\ref{eq:V1function}), we can easily see for $\tau > 1$ that
\begin{align}
\frac{\tau P_1 \gamma}{N_0 B + \mu \tau P_2} = \frac{P_1 \gamma}{\frac{N_0 B}{\tau} + \mu P_2} > \frac{P_1 \gamma}{N_0 B + \mu P_2}, \label{eq:tau-equation}
\end{align}
which leads to the result that
\begin{align}
V_1(\tau P_1, \tau P_2, B) > V_1(P_1, P_2, B). \label{eq:tau-equation2}
\end{align}
Hence, for given $P_1 < P^{\max}$ and $P_2 < P^{\max}$, we can increase the value of $V_1$ by increasing the power values to $\tau P_1$ and $\tau P_2$ for some $\tau > 1$ (with which the maximum power constraint $P^{\max}$ is still satisfied). Therefore, with this characterization, we conclude that in order to achieve the maximum value of $V_1$, we should have $P_1$ or $P_2$ attain its maximum value.

Next, we prove that $V_1$ is an increasing function of bandwidth $B$. Let us define $\chi = e^{-\theta T_c B\log\left(1+\frac{P_1 \gamma}{N_0 B +\mu P_2}\right)}$. Taking the first derivative of $V_1(P_1, P_2, B)$ with respect to $B$, we obtain
\begin{small}
\begin{align}
\hspace{-.3cm} \frac{\partial V_1}{\partial B} = \frac{\theta T_c \mathbb{E}_{\gamma}\left\{\chi \big(\ln{(1 + \frac{P_1 \gamma}{N_0 B +\mu P_2})} - \frac{P_1 \gamma_k N_0 B}{(N_0 B +\mu P_2 + P_1 \gamma)(N_0 B +\mu P_2)}\big)\right\}}{(\mathbb{E}_{\gamma}\{\chi\})^2\ln2}. \label{der_V_B_1st}
\end{align}
\end{small}
\normalsize
Let us also define
\begin{gather}
g(x) = \ln\left(1 + \frac{1}{x}\right) - \frac{1}{1+x}.
\end{gather}
The first derivative of $g(x)$ with respect to $x$ is
\begin{align}
\frac{d g(x)}{d x} = -\frac{1}{x(1+x)^2} < 0,
\end{align}
and hence $g(\cdot)$ is a decreasing function of $x \ge 0$. Moreover, $\lim_{x \to 0} g(x) = \infty$ and $\lim_{x \to \infty} g(x) = 0$. Thus, $g(x) \geq 0$ for all $x \ge 0$, which also implies that
\begin{gather}
\ln\left(1 + \frac{1}{x}\right) \ge \frac{1}{1+x} \quad \text{ for } x \ge 0. \label{eq:log_lowerbound0}
\end{gather}

Now, assume $x = \frac{N_0 B + \mu P_2}{P_1 \gamma}$. Then, we have
\begin{align}
\ln{\left(1 + \frac{P_1 \gamma}{N_0 B +\mu P_2}\right)} & \geq \frac{P_1 \gamma}{N_0 B +\mu P_2 + P_1 \gamma} \label{eq:log_lowerbound1} \\
& > \frac{P_1 \gamma}{N_0 B +\mu P_2 + P_1 \gamma}\frac{N_0 B}{N_0 B +\mu P_2}, \label{eq:log_lowerbound2}
\end{align}
where (\ref{eq:log_lowerbound1}) follows from (\ref{eq:log_lowerbound0}), and (\ref{eq:log_lowerbound2}) is due to the fact that $\frac{N_0 B}{N_0 B +\mu P_2} \le 1$.
The lower bound in (\ref{eq:log_lowerbound2}) shows that the derivative in (\ref{der_V_B_1st}) is greater than zero because the numerator is greater than zero. Therefore, we conclude that $V_1$ is an increasing function of $B$.

Note that these derivations immediately apply to
\begin{align} \label{eq:V2function}
V_2(P_1, P_2, B) = \left(\mathbb{E}_{\gamma}\left\{ e^{-\theta B T_c \log\left(1+\frac{P_2 \gamma}{N_0 B + \mu P_1}\right)}\right\}\right)^{-1}
\end{align}
due to the symmetry and similarity in the formulations.

Finally, we consider two target values $V_1^*$ and $V_2^*$ for the functions $V_1$ and $V_2$, respectively, i.e., $V_1(P_1, P_2, B) = V_1^*$ and $V_2(P_1, P_2, B) = V_2^*$, and show that the minimum bandwidth $B$ required to achieve these target values is attained if $P_1 = P^{\max}$ or $P_2 = P^{\max}$. Assume that both power values are strictly less than the maximum level, i.e., $P_1 < P^{\max}$ and $P_2 < P^{\max}$, and $B_a$ is the bandwidth value with which we satisfy $V_1(P_1, P_2, B_a) = V_1^*$ and $V_2(P_1, P_2, B_a) = V_2^*$.  Then, as also discussed above, there exists $\tau > 1$ such that $P_{1a} = \tau P_1 \leq P^{\max}$ and $P_{2a} = \tau P_2 \leq P^{\max}$. With these increased power levels, we now have
$V1(P_{1a}, P_{2a}, B_a) > V1^*$ and $V2(P_{1a}, P_{2a}, B_a) > V2^*$ as shown in (\ref{eq:tau-equation}) and (\ref{eq:tau-equation2}). Since $V_1$ and $V_2$ are increasing functions of $B$, there exists $B_b < B_a$, such that $V_1(P_{1a}, P_{2a}, B_b) = V_1^*$ and $V_2(P_{1a}, P_{2a}, B_b) = V_2^*$. Therefore, if both $P_1 < P^{\max}$ and $P_2 < P^{\max}$, we can always increase the power values and lower the bandwidth requirement while attaining the target levels $V_1^*$ and $V_2^*$. Hence, the minimum required bandwidth is achieved if $P_1 = P^{\max}$ or $P_2 = P^{\max}$.


\subsection{Proof of the Required Conditions for Obtaining the Upper Bound $\partial^+\mathcal{G}$} \label{Proof:2}
Assume that there exists an upper boundary point $\mathbf{V}^u$ such that $\sum_{k=1}^{K}B_k < B$, i.e.,  $V_{(i-1)K+k}(P_{1,k}, P_{2,k}, B_k) = V_{(i-1)K+k}^u$ for all $i \in \mathcal{I}$ and $k \in \mathcal{K}$. From Theorem \ref{theo:min_bandw} and its proof in Appendix \ref{Proof:1}, we know that $V_{(i-1)K+k}$ is an increasing function of $B_k$. Then, there exists a small positive $\delta$ such that $\sum_{k=1}^{K}(B_k+\delta) < B$ and $V_{(i-1)K+k}(P_{1,k}, P_{2,k}, B_k+\delta) > V_{(i-1)K+k}^u$, which implies that $\mathbf{V}^u$ is not a upper boundary point. Similarly, assume that there exists a upper boundary point $\mathbf{V}^u$ such that $P_{1,k} < P_{1,k}^{max}$ and $P_{2,k} < P_{2,k}^{max}$ for some $k \in \mathcal{K}$. Again, from the proof in Appendix \ref{Proof:1}, we know that we can find a $\tau > 1$ such that $\tau P_{1,k} < P_{1,k}^{max}$ and $\tau P_{2,k} < P_{2,k}^{max}$, and with these increased power values, we have $V_{(i-1)K+k}(\tau P_{1,k}, \tau P_{2,k}, B_k) > V_{(i-1)K+k}^u$ . This also means that $\mathbf{V}^u$ is not a upper boundary point. Therefore, the upper boundary point $\mathbf{V}^u$ only occurs when $\sum_{k=1}^{K}B_k = B$ and at least one power value is at its maximum level, i.e., $P_{1,k} = P_{1,k}^{max}$ or $P_{2,k} = P_{2,k}^{max}$, for all $k \in \mathcal{K}$.

\bibliographystyle{IEEEtran}
\bibliography{FullDuplex_Monotonic}

\end{document}